\documentclass{nature}

\usepackage[utf8]{inputenc}
\usepackage{graphicx}
\usepackage[small, bf, normal]{caption}

\usepackage{amssymb}
\usepackage{amsmath}
\usepackage{mathrsfs}
\usepackage{color}
\usepackage{times}
\usepackage{float}
\usepackage{rotating}
\usepackage{epstopdf}
\usepackage{multirow}
\usepackage{pdflscape}

\usepackage[]{lineno}
\def\apj{Astrophys. J.}
\def\mnras{Mon. Not. R. Astron. Soc.}
\def\aap{Astron. Astrophys.}
\def\apjl{Astrophys. J. Let.}

\def\nat{Nature}

\def\physrep{Phys. Rep.}
 
\def\na{New Astron.}
\def\jcap{J. Cos. Astropar. Phy.}

\title{Universal structure of dark matter haloes
over a mass range of 20 orders of magnitude }
\author{Wang, J.$^{1,5}$\thanks{E-mail: jie.wang@nao.cas.cn}, Bose, S.$^2$,  Frenk, C. S.$^3$\thanks{E-mail: c.s.frenk@durham.ac.uk}  Gao, L.$^{1,5}$, Jenkins, A.$^3$, Springel, V.$^4$ \& White, S. D. M.$^4$\thanks{E-mail: swhite@MPA-Garching.mpg.de}}

\begin{document}
\setlength{\unitlength}{1mm}
\sloppy

\maketitle

\begin{affiliations}
\linespread{1.0}\selectfont{}
  \item Key Laboratory for Computational Astrophysics,National Astronomical Observatories, Chinese Academy of Sciences, 20A Datun Road, Beijing 100101, China
  \item Center for Astrophysics $\vert$ Harvard \& Smithsonian, 60 Garden Street, Cambridge, MA 02138, USA
  \item Institute for Computational Cosmology, Department of Physics, Durham University, South Road, Durham, DH1 3LE, UK
  \item Max-Planck-Institut für Astrophysik, Karl-Schwarzschild-Str. 1, 85748 Garching, Germany
  \item School of Astronomy and Space Science, University of Chinese Academy of Sciences, Beijing 100039, China
\end{affiliations}

\begin{abstract}

  Cosmological models in which the dark matter consists of cold
  elementary particles predict that the halo population should extend
  to masses many orders of magnitude below those where galaxies can
  form \cite{Bluementhal_84,White_91,Bertone2005}.  Using a multi-zoom 
  technique, we report a consistent
  cosmological simulation of the formation of present-day haloes over
  the full mass range populated when dark matter is assumed to be a
  Weakly Interacting Massive Particle (WIMP) of mass
  $\sim$100~GeV. The simulation has a dynamic range of 30 orders of
  magnitude in mass and resolves the internal structure of hundreds of
  Earth-mass haloes in as much detail as that of hundreds of rich
  galaxy clusters.  We find that halo density profiles are universal
  over the entire mass range and are well described by simple
  two-parameter fitting formulae \cite{NFW_97,Navarro_04}. Halo mass and concentration are
  tightly related in a way that depends on cosmology and on the nature
  of the dark matter. At fixed mass, concentration is independent of
  local environment for haloes less massive than those of typical
  galaxies.  Haloes over the mass range
  $(10^{-3}-10^{11})~M_\odot$ contribute about equally (per logarithmic
  interval) to the dark matter annihilation luminosity, which we find to be smaller than all previous estimates by factors ranging up to one thousand\cite{Bertone2005}.
\end{abstract}

Figure~\ref{fig:Fig1} illustrates our simulation scheme. The top left
panel shows the present-day distribution of dark matter in a slab cut
from a large cosmological simulation (L0) identical to the 2005
Millennium Simulation\cite{Springel_05}, except that cosmological
parameters are updated to reflect recent analyses of CMB data from the
Planck satellite. The total mass in this simulation is about
$10^{19}\,{\rm M}_\odot$. The circle outlines a spherical region chosen to
avoid any of the more massive structures.  The material in this region
was traced back to the initial time and used to define a Lagrangian
volume within which the particle count was increased by a factor of
about 2000, the particle mass was decreased by the same factor, and
the representation of the linear cosmological fluctuation field was
extended to $\sim 10$ times smaller scale while retaining all structure
present in the original simulation. The mass outside this ``zoomed''
region was consolidated into a smaller number of particles whose
individual mass increases with distance from its centre. These new
initial conditions were then integrated down to the present day. The
top middle panel of Fig.~\ref{fig:Fig1} shows a projection of the mass
within the largest sphere enclosed in the high-resolution region. It
has resolution 2000 times better in mass and $\sim 10$ times better in
length than the first panel, but contains a comparable number of well
resolved haloes (i.e~made up of $10^4$ or more simulation particles).

The small circle in this panel outlines a spherical subregion of this
first level zoom (L1) which avoids any larger structures. It was again
traced back to the initial conditions, refined by another factor of
500 in mass, and resimulated to give a second level zoom (L2) for
which the final structure within the high-resolution region is shown
in the top right panel of Fig.~\ref{fig:Fig1}.  This whole process was
repeated eight times, each revealing ever smaller structures, to
give a final simulation (L8c) with eight levels of refinement and a
high-resolution particle mass of $\sim 10^{-11}\, {\rm M}_\odot$,
hence a dynamic range of 30 orders of magnitude.  The final mass
distributions in the high resolution regions at each stage are shown
in the remaining panels of Fig.~\ref{fig:Fig1}.  Their initial
conditions were set using second-order Lagrangian perturbation theory
with an initial power spectrum with Planck parameters together with a
free-streaming cutoff at small spatial scales corresponding to a
thermal WIMP, which, for illustrative purposes, we assume to have mass
100~GeV. One of the zooms (L7c) was repeated without this cutoff
(giving L7) in order to understand its effects on halo structure (see
the methods section for further details).

Considerable effort was needed to ensure that the initial conditions
procedure, the force calculation accuracy and the time integration
scheme of the simulation code were adequate to give reliable results
over such a large dynamic range. In the methods section we describe
some of these improvements, and we present
convergence tests that demonstrate that they were successful. The more
massive haloes in the high resolution region at each level can all be
individually identified in the parent level, making it possible to
check that the masses agree in the two cases. For the most massive
haloes, the resolution of the parent level is sufficient to test that
their radial density profiles also agree. The plots in the methods section
show that both these tests are passed for all adjacent
level pairs, giving us confidence that our results for the internal
structure of dark matter haloes are reliable for
$10^{-6}< M_{\rm halo}/{\rm M}_\odot < 10^{15}$, the entire halo mass range
that should be populated if the dark matter is a 100 GeV WIMP.

Figure~\ref{fig:Fig2} shows the first major result of this article. At
each level of our simulation we identify a sample of a few tens of
well resolved, quasi-equilibrium haloes of similar mass. For these, we
construct a mean, spherically averaged mass density profile which we
compare with two well known two-parameter fitting formulae, the NFW
profile\cite{NFW_96},
\begin{equation}
\rho(r) = \rho_sr_s^3/r(r+r_s)^2,
\label{eqn:NFW}
\end{equation} 
where $\rho_s$ and $r_s$ are the characteristic density and scale
radius respectively, 
and the Einasto profile\cite{Navarro_04,Einasto_65},
\begin{equation}
\rho(r) = \rho_{-2}\exp [ -2\alpha^{-1}((r/r_{-2})^\alpha -1 )],
\label{eqn:Einasto}
\end{equation} 
where $r_{-2}$ is the radius at which the logarithmic slope is $-2$,
and $\alpha$ is a shape parameter which we fix to $\alpha =
0.16$. These formulae were fit to the mean profiles at each level over
the radial range where these are numerically
robust. Fig.~\ref{fig:Fig2} shows differences between the measured
profiles and these best fits in two different ways. The upper panel
gives the logarithmic slope of the profiles as a function of
$r/r_{-2}$, where $r_{-2}$ is the characteristic radius of the best
Einasto fit. In such a plot, each fitting formula predicts a universal
curve, a Z-shaped transition between values of $-1$ and $-3$ in the
NFW case, and a smoother, more gradual change of slope in the Einasto
case. Over 20 orders of magnitude in mass, the mean profiles of the
simulated haloes are all very similar and are closer to the latter
case than to the former. The only clear exception is that the curve for
L0, representing haloes of moderately rich galaxy clusters, is noticeably steeper than the
others. Larger values of $\alpha$ have previously been shown to give a
better fit to such objects, but the trend in $\alpha$ does not
continue to the much lower masses we have now simulated. The lower
panels show that over the factor of about $10^4$ in density for which
these profiles are robustly measured, NFW fits are almost everywhere
accurate to better than about 10\% and Einasto fits to a few
percent. This universality over 20 orders of magnitude in halo mass is
remarkable, not least because reliable simulation data at $z=0$  have not
previously been available for most of this range.

The mass of a dark matter halo is conventionally taken as that within
the virial radius, defined here as $r_{200}$, the radius enclosing a
mean density 200 times the critical value. Mass and concentration,
$c = r_{200}/r_{\rm ch}$, can then be used as alternative parameters
for the above fitting functions, with $r_{\rm ch} = r_s$ and $r_{-2}$,
respectively, for the NFW and Einasto cases. Fig.~\ref{fig:Fig3} shows
the mass-concentration relation in our simulation, considering only
haloes with enough particles for a reliable concentration
measurement ($>10^4$ at the higher levels, somewhat fewer in L0, L1
and L2). Each coloured band gives the [10\%, 90\%] range for haloes at
a given level, with a white line indicating the median concentration
at each mass. Over the mass range
$10^{15} > M_{200}/{\rm M}_\odot > 10^{10}$ relevant for galaxy
clusters and for all but the very faintest galaxies, concentration
rises quite rapidly with decreasing mass. The relation becomes
shallower for lower mass haloes, however, and eventually turns down as the
free-streaming mass is approached. This turn-down is most clearly seen
by comparing results for L7c and L8c, where the initial conditions 
included a free-streaming cutoff, with
those for L7, where they did not. In the methods section we compare
matched objects in L7 and L7c, showing that the cutoff reduces the
concentration of individual haloes by an increasing amount as the
free-streaming mass (about Earth mass for a 100~GeV WIMP) is
approached. Like all N-body simulations of structure
  formation with a free-streaming cutoff, both L7c and L8c form
  spurious small-scale clumps. As discussed in the Methods section,
  this does not affect the results of this article.

Other points of interest in Fig.~\ref{fig:Fig3} are that the scatter in concentration depends very little on halo mass, being about 0.15
dex over the full halo mass range plotted, and that previously
published mass concentration relations, while agreeing roughly for
galaxy- and cluster-mass haloes, give wildly divergent results when
extrapolated down to the halo masses which are simulated here for the
first time. Only one model \cite{Ludlow_14,Ludlow_16} represents our results relatively well, both with and without a free-streaming
cutoff. In the methods section we give a simpler fitting formula which fits our numerical data even better and follows their approach to predict the effects of varying the free-streaming scale. 

The concentration-mass relation is of critical importance for predicting WIMP annihilation radiation signals. Previous work implied that these should be dominated by haloes with mass relatively close to the free-streaming limit, but this changes substantially for the reduced concentrations we find (see the methods section for details). Structures down to very small scales should also be present in the outer regions of much more massive haloes, resulting in a substantial boost in the total amount (and a flattening of the radial profile) of their annihilation luminosity. Our simulation cannot address these issues directly, but it can be used to inform the further modelling required \cite{Springel_08a}.

The high resolution region of L8c is only about 300 pc across at the
final time and contains a total mass which is only about 1\% that of
the Sun, implying a mean density about 0.3\% that of L0. This low
value is a consequence of repeatedly choosing to refine regions that
avoid any massive nonlinear structure. It is still somewhat larger
than the median $z=0$ density of a universe with Planck cosmology
dominated by a 100~GeV WIMP\cite{Stuecker2018}.  One may nevertheless
question whether the haloes we have simulated can be considered
representative of the general population of similar mass objects. In
the methods section we test this by investigating how the
concentration of our haloes depends on the density of their immediate
environment, measured in a spherical shell between 5 and
$10\,r_{200}$. Remarkably, despite the low mean density of the higher
refinement levels, the distribution of this environment density is
centred just below the cosmic mean for all haloes less massive than
about $10^{10}\,{\rm M}_\odot$, with a spread of at least an order of
magnitude. In addition, such haloes show no systematic trend of
concentration with local density.  This encourages us to believe that
the concentration-mass relation of Fig.~\ref{fig:Fig3} should be
representative of the full halo population. Previous
attempts\cite{Diemand2005,Ishiyama2010,Anderhalden2013,Angulo2017} to
simulate the structure of very low-mass haloes have failed precisely
because they did not take account of the low-density larger scale
environment in which such haloes live at~$z=0$.

A final related issue is that our simulation follows dark matter only,
neglecting the effects of the 16\% of cosmic matter which is
baryonic. Both relative velocity and pressure
effects\cite{Tseliakhovich2010} are expected to prevent the gas from following the
dark matter on the very small scales we have simulated. While accurate
treatment of these effects is beyond present capabilities, given the
dynamic range we are considering, we may expect that at the higher
refinement levels they would increase the mean density (because on
average the baryons will be less underdense than the dark matter) but
reduce the growth rate of haloes (because this is driven by the dark
matter density only, rather than by the total density). Given that
halo concentration depends weakly on halo mass and not at all on local
environment density, we expect these effects to shift our results by
at most small factors, but this will require further work for
confirmation.

The universal halo structure we have demonstrated across 20 orders of magnitude in halo mass and the associated mass-concentration relation differ substantially from previously proposed extrapolations. This  affects predictions not only for annihilation signals, which depend strongly on the concentration of the lowest mass haloes, but also for perturbations of image structure in strong gravitational lenses and for  structure in stellar streams in the Galaxy's halo, both of which aim to constrain the nature of dark matter using haloes of mass $10^6$ to $10^9{\rm M}_\odot$.

\expandafter\ifx\csname url\endcsname\relax
  \def\url#1{\texttt{#1}}\fi
\expandafter\ifx\csname urlprefix\endcsname\relax\def\urlprefix{URL }\fi
\providecommand{\bibinfo}[2]{#2}
\providecommand{\eprint}[2][]{\url{#2}}
\setlength{\itemsep}{0ex}

\clearpage
\section*{References}

\renewcommand{\refname}{\vspace{-0.5cm}}
\bibliographystyle{naturemag}

\begin{addendum}

\item[Acknowledgements] 
This work was supported by the National Natural Science Foundation of China (NSFC) under grant 11988101,the National Key Program for Science and Technology Research Development (2017YFB0203300), and the UK Science and Technology Facilities Council (STFC) consolidated grant ST/P000541/1 to Durham. JW and LG acknowledges support by the NSFC grant 11373029, 11873051 and 11851301. SB acknowledges support from Harvard University through the ITC Fellowship. CSF acknowledges support by the European Research Council (ERC) through Advanced Investigator grant DMIDAS (GA 786910). The visualisations in Fig. 1 were produced using the py-sphviewer code (github.com/alejandrobll/py-sphviewer). This work used the DiRAC@Durham facility managed by the Institute for Computational Cosmology on behalf of the STFC DiRAC HPC Facility (www.dirac.ac.uk). The equipment was funded by BEIS capital funding via STFC capital grants ST/K00042X/1, ST/P002293/1 and ST/R002371/1, Durham University and STFC operations grant ST/R000832/1. DiRAC is part of the National e-Infrastructure.

\item[Author contributions] 

JW: contributed to design of project, carried out simulations, performed analysis, wrote first draft of paper

SB: contributed to design of project, carried out simulations, performed analysis

CSF: contributed to concept of project and its design, analysis of results and writing of final draft

ARJ: contributed to design of project, upgraded initial conditions code and generated initial conditions, contributed to analysis of results

GL: contributed to design of project, analysis of results 

VS: wrote and upgraded simulation code, carried out simulations, contributed to analysis of results

SDMW: initially conceived the project, contributed to its design, analysis of results and writing of final draft 

\end{addendum}

\clearpage

\renewcommand{\figurename}{Figure}

\begin{figure}
\begin{center}
\vspace{-1cm}
\includegraphics[width=0.75\textwidth,angle=0]{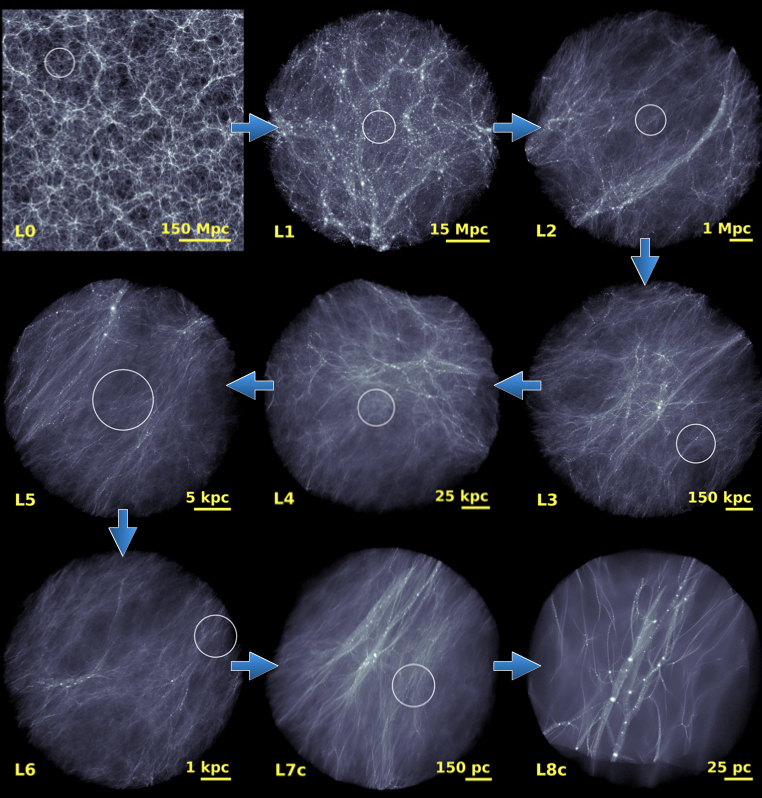}
\end{center}
\caption{\linespread{1.3}\rm \textbf{Projected dark matter density maps at each simulation 
    level.} Images of the dark matter distribution in a
  slice 30 Mpc thick through the base level of our simulation (L0) and
  in spheres almost entirely contained within the higher resolution
  region of each of the eight successive levels of zoom (L1 to
  L8c). The zoom sequence is indicated by arrows between the panels,
  and a circle in each of the first eight panels indicates the zoom
  region shown in the next panel. Bars give a length scale for each
  plot. In the first panel the largest haloes have a mass similar to
  that of a rich galaxy cluster, whereas in the last panel the
  smallest clearly visible haloes have a mass comparable to that of
  the Earth.}
\label{fig:Fig1}
\end{figure}
\clearpage

\begin{figure}
\begin{center}
\vspace{-2cm}
\includegraphics[width=0.65\textwidth,angle=0,trim=0.1cm 1cm 1cm 6cm,clip=true]{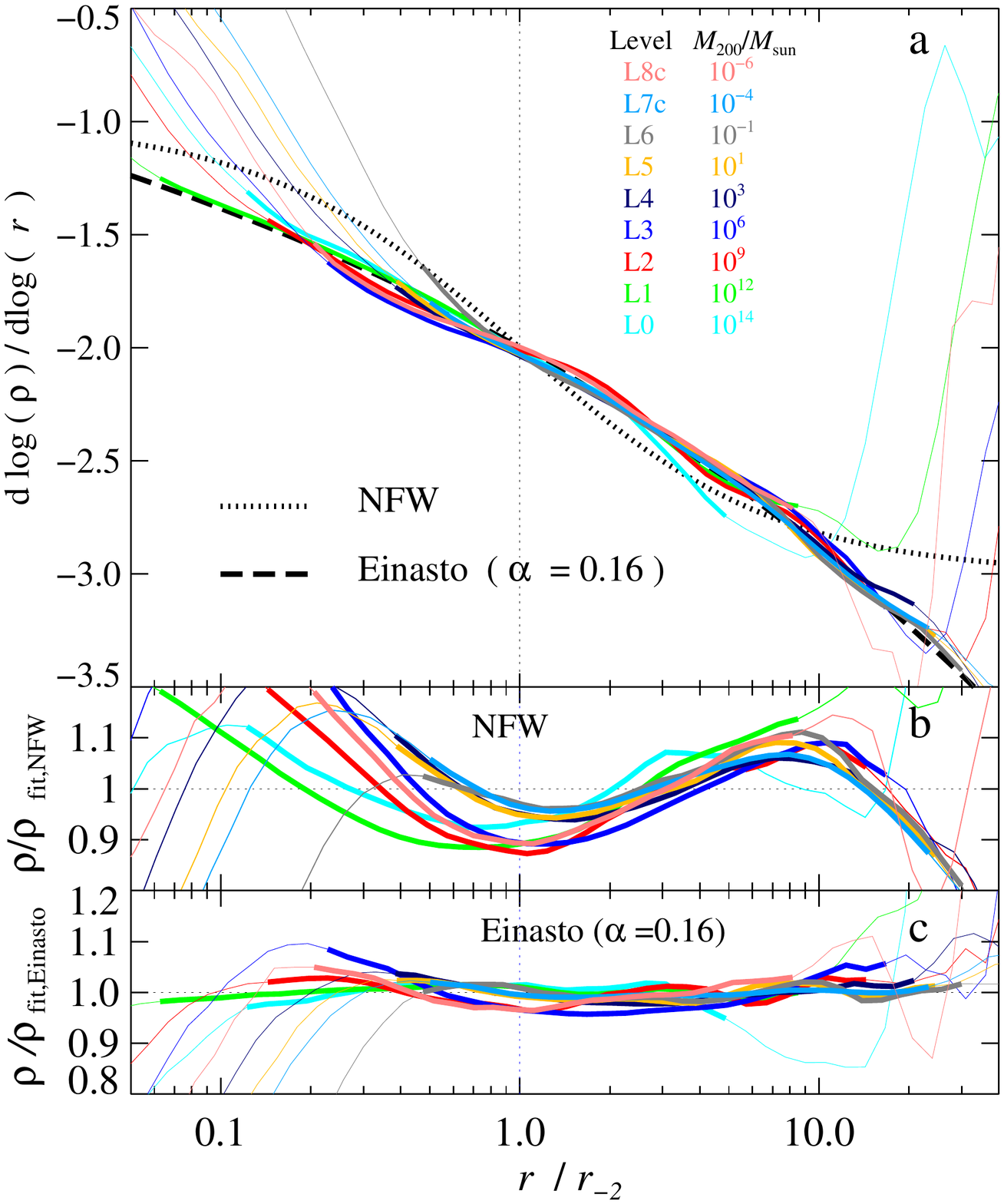}
\end{center}
\vspace{-1.4cm} 
\caption{\linespread{1.3}\rm \textbf{Density profiles for haloes with mass between that of the Earth and that of a rich galaxy cluster.}
    As described in detail in the methods section,
  results for all well-resolved equilibrium haloes in a narrow mass
  bin at each level are averaged together. {\it Panel a}: the
  logarithmic slope ${\rm d}\log(\rho)/{\rm d}\log(r)$ is shown as a function of
  radius normalised by $r_{-2}$. The result for each level is
  represented by a different colour, as indicated in the legend. A
  thicker line is used over the most reliable range between the
  convergence radius $r_{\rm conv}$ and $r_{200}$. The number of
  haloes in each stack is listed in Extended Data
  Table~\ref{tab:tab1}. Predictions for NFW and Einasto profiles are
  shown as dotted and dashed black curves, respectively. {\it Panel b}: the ratio of each stacked profile to the best fit NFW
  profile is shown as a function of $r/r_{-2}$. {\it Panel c}:
  the same but for the Einasto profile (with $\alpha$ fixed at 0.16).
}
\label{fig:Fig2}
\end{figure}
\clearpage

\begin{figure}
\begin{center}
\vspace{-4cm}
\includegraphics[width=0.65\textwidth,angle=0,scale=1.1,trim=1cm 3.5cm 0cm 5cm,clip=true]{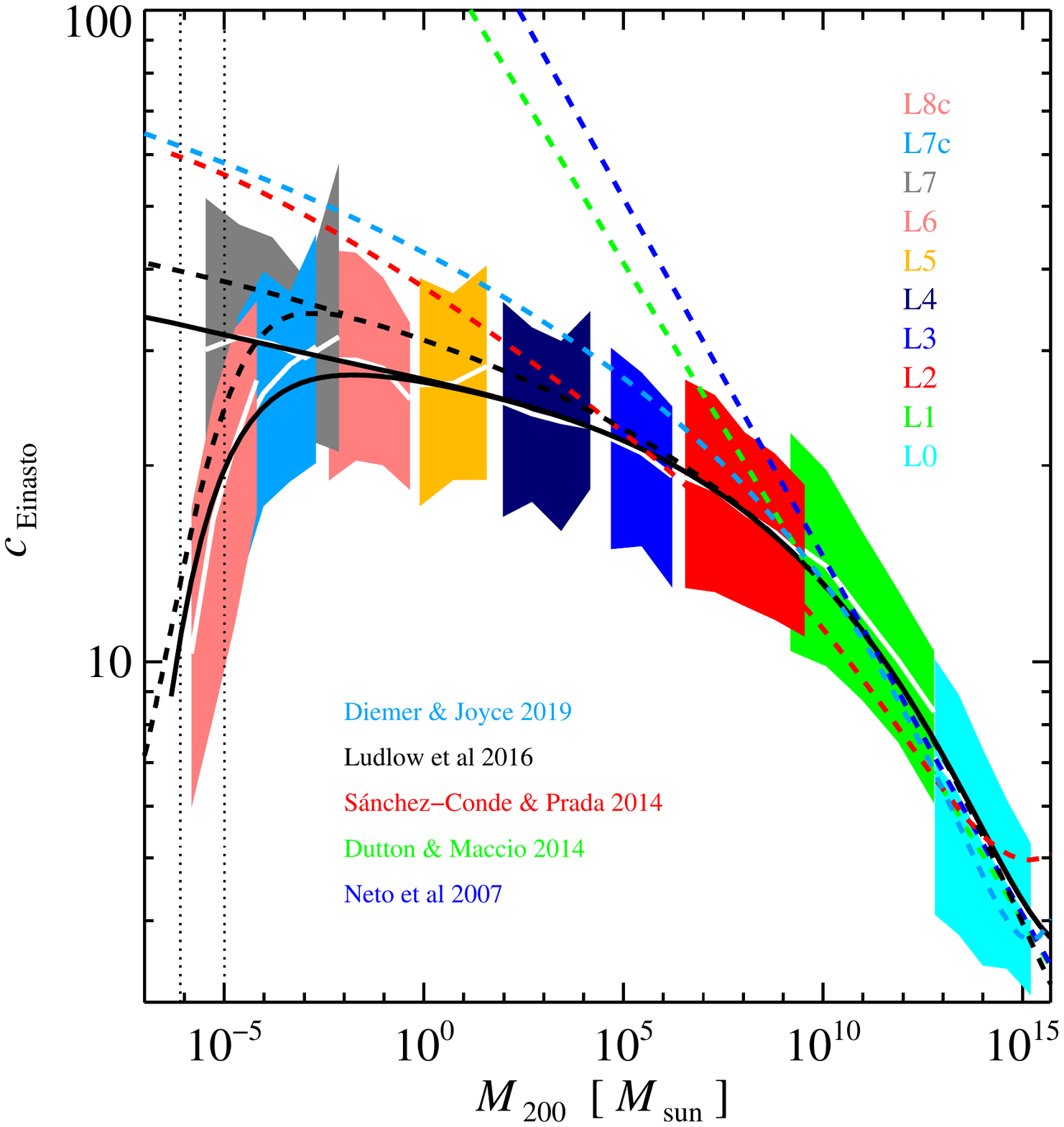}
\end{center}
\vspace{-2cm}
\caption{\linespread{1.3}\rm \textbf{Halo concentration as a
    function of mass over a mass range of 20 orders of magnitude.}
  The median values of the concentration,
  $c _{\rm Einasto}= r_{200}/r_{-2}$ (from the best-fitting Einasto
  profile), in each mass bin are shown as white curves, with coloured
  regions showing the {\it rms} scatter. As before, each zoom level is
  shown with a different colour, and we give results both for L7c,
  which has a free-streaming cutoff, and for the otherwise identical
  L7, which does not. Mass-concentration relations from five published
  models \cite{Neto_07,Dutton_14,Sanchez-Conde_14,Ludlow_16,Diemer_19} are shown as smooth dashed lines in different colours. Solid black lines show the fitting
  formulae given in the Methods section.
  The pairs of solid and dashed black lines give predictions 
  for the cases with and without
  a free-streaming cutoff.  The vertical dotted lines
  indicate the limits below which spurious haloes are expected to
  affect L7c and L8c (see Methods). }
\label{fig:Fig3}
\end{figure}
\clearpage

\newpage

\begin{methods}
\subsection{Simulations.} 
\label{simulations}
The hierarchical resimulation strategy that allows us to follow the
evolution of haloes over 20 orders of magnitude in mass was described
in the main body of this article. The base level (Level 0 or L0)
is a cube of length $738\, \rm Mpc$ and particle mass
$1.55\times 10^9\, {\rm M}_{\odot}$. At subsequent levels (L1-L8c) the mass
resolution increases by factors between a few hundred and 2000 and the volume decreases by
similar factors until the particle mass reaches
$1.6 \times 10^{-11}\, {\rm M}_\odot$ in L8c. At each level, well-resolved
haloes (i.e.~with $>10^4$ particles within the virial radius) span $2$-$3$
orders of magnitude in mass, ranging from
$M_{200} = 10^{15}\, {\rm M}_\odot$ in L0 to $M_{200}=10^{-6}\, {\rm M}_{\odot}$ (the
Earth's mass) in L8c.  Here $M_{200}$ is defined as the mass within a sphere enclosing a  mean density 200 times the
critical value. The parameters of the various levels of our simulation are listed in
Extended Data Table~\ref{tab:tab1}.
The simulation assumes a $\Lambda$CDM cosmology with  Planck 2014
parameters\cite{Planck_14}. Specifically,
the mean matter density, mean baryon density and cosmological
constant, in units of the critical density, have values
$\Omega_m=0.307$, $\Omega_b=0.048$ and $\Omega_{\Lambda}=0.693$; the
present-day Hubble parameter is $H_0=67.77 {\rm km~s^{-1}~Mpc^{-1}}$; the power-law
  index of the power spectrum of primordial adiabatic perturbations is
  $n_s=0.961$; and the normalization of the linear power spectrum is
  $\sigma_8=0.829$.

\subsection{Linear Power spectrum.}

To create the displacement field for the initial conditions a linear
power spectrum is required which covers more than nine orders of
magnitude from the fundamental modes of the L0 volume to the Nyquist
cutoff of the L8c simulation.  For relatively large scales we use the same
linear power spectrum as the EAGLE
project\cite{Schaye_15}. This was computed using
the CAMB code\cite{Lewis_00} assuming Cold Dark Matter with the values of the cosmological
parameters given previously. Since we are only modelling the dominant dark matter component, we use
the BBKS fitting formula (Eqn~G3) to extrapolate the power spectrum to very small scales.\cite{BBKS}  We adopt a similar approach to previous work\cite{Springel_08a,Springel_08b}, creating a
composite matter power spectrum that smoothly transitions from the
EAGLE matter power spectrum to the BBKS form
over wavenumbers $7 - 70\,{\rm Mpc}^{-1}$.
We determined by inspection that setting the parameter $\Gamma = 0.1673$ in the BBKS transfer function and adopting an effective normalisation, $\sigma_8 = 0.8811$, matches the BBKS power spectrum accurately to the EAGLE power spectrum over the transition wavenumber range; we interpolate linearly in log wavenumber over this range to produce a smooth power spectrum for pure CDM.

We then used the formulae of Green et al.\cite{Green_04} to represent the free-streaming cut-off
expected for a 100GeV WIMP. This corresponds to a specific particle physics model, but the corresponding mass scale is close to the peak of the posterior probability for CMSSM space\cite{Martinez_09} and can still be considered representative in view of more recent constraints.\cite{Bertone_18} We show below how to adapt our results to alternative particle physics models that would predict a free-streaming cut-off on a different scale.

\subsection{Making the initial conditions.}

While the setting up and running of zoom simulations has become
commonplace in the field of numerical cosmology, the initial
conditions required for the present project are much more extreme in
terms of the range of mass and length scales modelled than in any
previously published simulation. These exceptional demands have driven
developments that go beyond the techniques described in previous
work\cite{Springel_08a,Jenkins2010,Jenkins2013}.

The initial conditions for levels L1 to L8c were created and evolved
sequentially in order of increasing mass resolution. After each level
was completed, a region avoiding any massive halo was selected from
its high resolution region and this then became the next level (see
Fig.~\ref{fig:Fig1}). The amplitude and phase of the initial
fluctuations present in the initial conditions of all lower levels
were retained, but the amplitude and phase of all higher frequency
fluctuations added at the new level were set independently and at
random according to the power spectrum. In principle, we could make
initial conditions at the resolution of L8c for any Lagrangian region
within L0 without running any intermediate levels, but in the great
majority of cases this would result in all of the mass being
incorporated into a single halo of mass larger than that of the entire
high-resolution region of L8c.

The specific features that emerge at any redshift,
for example, the positions, masses and orientations of individual
haloes or filaments, are a consequence of our particular realisation
of the linear initial conditions, i.e. of our adopted power spectrum
together with the specific phases and amplitudes chosen for each wave
in a Fourier space representation of the initial Gaussian random
field. Our phase information was taken from the Panphasia white noise
field\cite{Jenkins2013, JenkinsBooth2013}, an extremely large single
realisation of a Gaussian white noise field with a hierarchical octree
structure.  Because the Panphasia field is completely specified ahead
of time, all of the structure uncovered at all resolutions is
essentially predetermined, as is the similar structure that would be
uncovered by a different hierarchical zoom into any other region of
the L0 cube.

The creation of initial conditions at each zoom level can be divided
into three stages: Stage~1 is to specify the region of interest;
Stage~2 is to build a particle load focussing most of the particles,
and therefore most of the computational effort, in the small region of
interest, while aggregating particles for lower levels so that the
computational time for these regions is reduced while maintaining
accurate tidal forces in the high-resolution region; Stage~3 is to
generate and apply the displacement field to the particle load, and
assign velocities to each particle.

For Stage~1 we start by selecting a spherical region of interest at
redshift zero from a previously completed simulation of the lower
level. For L1, this is the cosmological simulation L0, but for all
higher levels it is itself a zoom simulation. The region was selected
by eye using projections of the density field to avoid large haloes
that were previously simulated with good resolution and would be
prohibitive to simulate at much higher resolution. At the same time we
avoided regions that were more underdense than necessary, as these
would yield few new haloes. The region size was dictated by the cost
of resimulating at the resolution desired for the next level, given
that we could afford simulations with a few billion high-resolution
particles.  Having selected a sphere, we then use its particles to
determine the location and shape of the corresponding Lagrangian
region by binning their high-redshift positions onto a $40^3$ cubic
grid just large enough to enclose them all.  Within this cube, we
define a simply connected region by selecting grid cells that either
contain a particle or are adjacent to one that does.

For constructing the particle load in Stage~2, we use a set of cubes
with a variety of sizes that tesselate the entire simulation volume.
In each cube we place one or more particles of identical mass in an
arrangement that ensures that the centre of mass of the particles
within every cube is at the cell centre, and we choose the total
particle mass in each cell so that it has precisely the mean density
of the universe. We also place the particles as evenly as possible
within each cube in the sense that if that cube were tesselated over
all space, the gravitational forces on each particle due to all other
particles would be essentially zero.  For the region outside the high
resolution cube we lay down a set of `tidal' particles arranged within
a series of concentric cubic shells centred on the high resolution
cube.  For the initial conditions of the highest level, L8c, more than
47 million tidal cells, each containing a single particle, are used to
represent the mass distribution in the lower levels.  

The particles
within the high-resolution region are the ones that form the
structures analysed at each level.  In L8c, for example, we place a
preprepared set of about 50000 particles with a glass-like structure
in each of the retained cells from our $40^3$ mesh.  This glass-like
arrangement is created in a small periodic cube and results in the net
gravitational force on every particle being extremely small. We also
enforce the condition that the centre of mass of the glass be exactly
at the cell centre. Because the glass is generated using periodic
boundary conditions, it is simple to tile the entire high-resolution
region with multiple replicas.  The number of particles in the glass
determines the mass resolution in the high resolution region.

In Stage~3, we generate and apply the displacement field following
exactly the method described in previous work\cite{Jenkins2013}. The
displacement field is computed using Fourier methods for a series of
concentric meshes centred on the high-resolution cube.
The top level mesh covers the entire domain and the smallest mesh just covers the high-resolution cube.  Each
successive mesh is exactly half the linear size of the one above, and
adds additional independent information taken from the Panphasia field
so as to be able to double the linear resolution of the displacement
field.  The L8c simulation required 23 levels in total with the
smallest mesh being approximately 180~pc on a side.

We have used second order Lagrangian perturbation theory (2lpt) to
create the displacement and velocity fields for most of the initial
conditions.  In practice, however, we have found that using first-order (Zeldovich) initial conditions instead of
2lpt makes no significant difference to the
results provided the starting redshift is high enough. In particular,
for our chosen starting redshift of 127 for levels 0-2, and 255 for
levels 3-8, there was no significant difference in the halo density
profile or its concentration between runs using the Zeldovich and 2lpt
initial conditions.  Nonetheless, we used the 2lpt initial conditions
for all but levels 4 and 6.

\subsection{Simulation code.}

The simulations were run with GADGET-4, a new version of the
well-tested GADGET\cite{Gadget2} cosmological N-body code. A number of
improvements were implemented in this code to allow the extreme zooms
considered here to be executed with the required accuracy.  The most
relevant is an extension of the hierarchical multipole force
computation algorithm to higher expansion order, yielding better force
accuracy for given computational cost. A further efficiency gain comes from replacing the one-sided Barnes \& Hut tree
algorithm\cite{Barnes1986} with a Fast Multipole
Method\cite{Dehnen2000} (FMM), where the multipole expansion is
carried out symmetrically both at the source and the sink side of two
interacting particle groups.

The extreme dynamic range of our zooms revealed two problems that had
not shown up in more conventional cosmological simulations with
uniform mass resolution. Because the magnitude of the peculiar
acceleration vector, $\textbf{a}$, of particles in the small
structures targeted here is typically dominated by matter
perturbations on much larger scales, the local timestep criterion most
commonly employed in cosmological N-body cold dark matter simulations,
$\Delta t \propto (\epsilon / |\textbf{a}|)^{1/2}$, where $\epsilon$
is the gravitational softening length, often fails to provide a
reasonable proxy for the local dynamical time in our smallest dark
matter halos. Rather, it tends to become unrealistically small because
$|\textbf{a}|$ remains at the large values characteristic of the
resolved cosmic large scale structures in our $738\,{\rm Mpc}$
periodic box, whereas $\epsilon$ shrinks to the tiny scales resolved
in our calculations. We address this problem by applying a
hierarchical time integration algorithm\cite{Pelupessy2012} in which
the Hamiltonian describing the system is recursively split into parts
that evolve sufficiently slowly to be treated with a relatively long
timestep, and faster parts that require shorter timesteps. This
procedure effectively decouples the small-scale dynamics from the
large-scale forces. The above canonical timestep criterion
then yields a reasonable timestep for the smallest forming structures once
it is applied (some steps down the hierarchy) only to the partial
accelerations created by the high-resolution region itself.

A more subtle issue that becomes apparent with our very high dynamic
range arises from the fact that force errors in our hierarchical
multipole algorithm are spatially correlated. As a result,
neighbouring particles normally have very similar node interaction
lists. Formally, this creates force discontinuities across boundaries
of the hierarchically nested cubes of the global octree geometry
because the interaction lists and the field expansions (in the case of
FMM) change there. Small haloes, for which internal peculiar
accelerations are small compared to that induced by large-scale
structure, can be significantly affected by such errors if they are
cut by an octree boundary corresponding to a geometrically large
node. In such cases, the force error discontinuity can be appreciable
relative to the peculiar acceleration. At high redshift this error can
build up over many timesteps if the halo is nearly at rest relative to
the octree pattern.  To alleviate such effects, we decorrelate these
errors in time by translating the whole particle set by a random
vector (drawn uniformly from the cubic volume) after every
timestep. Physically, this does not change anything as the periodic
system is translationally invariant. Numerically, it causes the above
errors to average out in time, thereby preventing the 
build-up of sizeable momentum errors over many steps.

\subsection{Convergence.}

A critical test of our numerical techniques is convergence in the
properties of our simulations. We first examine maps of
the mass distribution in common regions of adjacent zoom levels. As an
example, in Extended Data Fig.~\ref{fig:FigA1} we compare projected
density distributions in L1, L3 and L8c with the corresponding
distributions in the same region of the parent level. It is clear that
large-scale structure in the simulations is converged.

We next check the convergence properties of the halo mass function,
again by comparing results for common regions of adjacent levels.
Mass functions of haloes in spherical volumes of radius approximately 90\% that of
the radius of the entire high-resolution region are shown
as solid curves in Extended Data Fig.~\ref{fig:FigA2}, with different
levels indicated by different colours.  The mass functions of haloes
in the same region in the parent simulations are shown as dotted
curves.

The convergence of the halo mass functions in adjacent zoom levels is
remarkable. Small differences appear at low masses when comparing
simulations which resolve the free-streaming cutoff (L7c and
L8c). These stem from the presence of spurious haloes that form due to
discreteness effects when a cutoff in the power spectrum is
resolved\cite{Wang_07}. The two vertical dotted lines indicate the masses below
which the abundance of these spurious haloes becomes
important in the high-resolution regions of levels L7c and L8c.\cite{Wang_07} For these
cases, convergence can be tested only to the right of the
dotted lines and, as the figure shows, in this regime convergence is
very good. In this article we have only considered halos in L7c and L8c
with mass above these limits.

A convergence test of the internal structure of halos is shown in
Extended Data Fig.~\ref{fig:FigA3}. Here we compare the density
profile of one of the most massive haloes in a given level (solid
lines) with the same halo in its parent level (dashed lines). The
profiles of haloes from the parent simulation are plotted as thick
solid lines in the radial range between the ``convergence radius''
\cite{Power_03} and $r_{\rm 200}$. The ratio of the density profiles
of the matched pairs is plotted in the bottom panel of the figure. At
radii larger than the convergence radius, the profiles agree to within
a few percent.

In summary, Extended Data Figs.~\ref{fig:FigA1}, \ref{fig:FigA2} and
\ref{fig:FigA3} show that in the regime where convergence can be
tested, the spatial distribution, the abundance and the density
profiles of haloes converge remarkably well over a factor of several
hundred in mass resolution for all adjacent pairs of levels in our simulation.

\subsection{Global properties of our simulation levels} 

Extended Data Table 1 lists a number of properties of the high resolution regions at each level of our simulation: $n_p$ is the number of high-resolution particles and
$m_p$ is the mass of each one, so that  $M_{\rm tot}= n_p m_p$ is the total mass of the high-resolution region. The {\it rms} linear fluctuation (extrapolated to $z=0$) expected in a spherical region which on average contains this mass is given as $\sigma(M_{\rm tot},z=0)$. For levels 2 and higher this number exceeds unity, reaching 17.6 in L8c. As a result, the actual $z=0$ densities of the high resolution regions, given as $\langle\rho\rangle/\rho_{\rm mean}$, are small, typically a few percent. This is expected since most of the mass of the universe is contained in ``high  mass" haloes that have, by construction, been excluded from the higher simulation levels.\cite{Stuecker2018}   The fact that typical present-day haloes of very low mass
(e.g.~Earth mass) form from Lagrangian regions with atypically low (linear) overdensity on larger scales (e.g.~$1 - 10^6\, {\rm M}_\odot$) explains why previous work has been unable to follow their evolution to low redshift.\cite{Angulo2017}

\subsection{Density profiles.}

We selected only ``relaxed'' or ``equilibrium'' haloes defined as
those which satisfy the following two criteria\cite{Neto_07}: {\em
  (i)} the mass fraction in subhaloes within the virial radius is less
than 0.1, and {\em (ii)} the offset between the centre of mass and the
minimum of the potential is less than 0.07$r_{200}$. As listed in Extended Data Table 1,
more than 90\% of our well-resolved haloes satisfy these conditions at all levels (more than 95\% at the highest levels). This can be understood as a result of the relatively high typical formation redshifts of the haloes (also listed in the Table) although these are considerably lower than found in earlier work which was unable to follow such low-mass haloes to 
low redshift. Interestingly, we find the lowest mass haloes to have {\it lower} typical formation redshifts than slightly more massive objects, consistent with the lower concentrations we find below. We note also that we found no higher mass particle within $0.95$ of the radius of the high-resolution region at any level, and that extremely few well-resolved haloes were contaminated by such particles. We excluded from our analysis any halo with such a particle within twice its virial radius, $r_{\rm 200}$. 

To make mean mass density profiles, we averaged binned mass
densities in the radial range $(0.001 - 10)\, r_{\rm 200}$ for haloes lying
between 0.8 and 1.2 times the central mass values listed in
Extended Data Table~\ref{tab:tab1}.  We then fitted NFW\cite{NFW_96} and
Einasto\cite{Navarro_04,Einasto_65} formulae (Eqns.~\ref{eqn:NFW}
and~\ref{eqn:Einasto}) to the stacked profiles using the bins between
the ``convergence'' radius\cite{Power_03} and $r_{200}$ by minimizing
the expression\cite{Neto_07}:
\begin{equation}
    \Psi^2=\frac{1}{N_{\rm bin}}\; \sum^{N_{\rm bin}}_{i=1}[\ln\rho_{\rm
      sim,i}-\ln\rho_{\rm fit,i}]^2 , 
\end{equation}
where $\rho_{\rm sim,i}$ and $\rho_{\rm fit,i}$ are the simulation
data and the fitted density profile in radial bin, $i$.  For the
Einasto fits the shape parameter, $\alpha$, was set to
0.16\cite{Gao_08} so that only two parameters are varied in both the
NFW and Einasto fits. 

The logarithmic slopes of the stacked radial density profiles of
haloes are plotted in Fig.~\ref{fig:Fig2} out to large radii,
$10\, r_{\rm 200}$. The ratios of the stacked profiles to the best-fit
NFW and Einasto formulae are plotted in the lower panels of this
figure and show that the NFW fits are almost everywhere accurate to better than 10\% and the
Einasto fits to a few percent. 

As may be seen in Fig.~\ref{fig:Fig2}, while Einasto profiles with $\alpha=0.16$
fit the data well overall, for haloes with $M_{200}>10^{12}{\rm M}_\odot$ the fits have relatively large residuals. We therefore carried out Einasto fits to
individual haloes with all three parameters free.  The resulting
median dependence of $\alpha$ on halo mass is well described by:
\begin{equation}
    \alpha=0.16+0.0238 \times (M_{\rm 200}/M_{*})^{\frac{1}{3}} 
\label{eqn:alpha_M200}
\end{equation}
where $M_{*}$ is defined by $\sigma(M_{*}) = 1.68$, where $\sigma(M)$
is the {\it rms} linear fluctuation within a sphere which on average
contains mass $M$. This extends previously published
formulae\cite{Gao_08} to much lower halo mass. For the Planck
cosmology we use here, $M_{*}= 1.14 \times 10^{14}\,{\rm M}_{\odot}$. We now
refit all halos using for each an $\alpha$ value given by
equation~\ref{eqn:alpha_M200} and adjusting only the two remaining
parameters. In this way, we obtain a robust estimate of the concentration-mass relation over the full halo mass range
accessed by our simulation.

The resulting relation between $M_{200}$ and
$c_{\rm Einasto}=r_{200}/r_{\rm -2}$ is shown in
Fig~\ref{fig:Fig3}. Simple extrapolations of empirical formulae
derived for halos of mass $M\geq 10^{10}\, {\rm M}_\odot$ overestimate the
concentrations of low-mass halos ($M_{200}< 10^6\, {\rm M}_\odot$) by large
factors. On the other hand, formulae derived from halo mass accretion
histories\cite{Ludlow_14,Ludlow_16} match our data better over
the entire halo mass range, both with and without a free-streaming
cutoff. We fit a simple parametrized form used
previously\cite{Lavalle_08,Sanchez-Conde_14} to the median
concentration-mass relation for levels L0 to L7 (i.e. with no
free-streaming cutoff) namely:
\begin{equation}
c_{\rm Einasto} (M_{\rm 200})= \sum^5_{i=0}c_{\rm i}  \left[\ln \frac{M_{\rm 200}}{h^{-1}{\rm M}_{\odot}}\right]^i 
\end{equation}

When the free-streaming cutoff is significant (i.e. for L7c and L8c),
the concentration drops exponentially at the low-mass end and the
relation is well fit by:
\begin{equation}
c_{\rm Einasto} (M_{\rm 200})={\rm exp} \left [ c_6 \times  \left(  {\frac{M_{\rm
      fs}}{M_{\rm 200}}}\right)^{\frac{1}{3}} \right] \times \sum^5_{i=0}c_{\rm i}
\left[\ln \frac{M_{\rm 200}}{h^{-1}{\rm M}_{\odot}}\right]^i. 
\end{equation}
In these relations the $c_i$ are dimensionless constants and the free-streaming
mass scale is given by
$M_{\rm fs}= \frac{4\pi}{3}\times ({\frac{2\pi}{k_{\rm fs}}}
)^3\times\rho_{\rm mean}$, where $k_{fs}$ is the
free-streaming wave-number defined by Equation~3 of Green et
al.\cite{Green_04}.  For a thermal WIMP of mass 100~GeV,
$M_{\rm fs} =7.3 \times 1 0^{-6}\, {\rm M}_\odot$ and $k_{fs} = 1.77\textrm{pc}^{-1}$. We find the following
best-fit values for the other parameters:
$c_i= [ 27.112, -0.381, -1.853 \times 10^{-3}, -4.141 \times 10^{-4},
-4.334 \times 10^{-6}, 3.208 \times 10^{-7}, -0.529 ] $ for
$i \in \left\{0, \ldots , 6 \right\}$.

\subsection{Environmental dependence.}
\label{Env-depen}

Our strategy for simulating haloes over the entire mass range expected
in a $\Lambda$CDM universe relies on successive resimulation of
low-density regions. An important question is then whether the
structure of these haloes is typical of the overall population. We can
address this by examining how the concentration of haloes of a
given mass varies with environment. We characterize the local  environment of
each halo by the mean density, $\left<\rho\right>$, averaged over a
surrounding shell with inner and outer radii 5 and 10 times the halo's
virial radius, $r_{200}$.

In Extended Data Fig.~\ref{fig:FigA4} we plot halo concentration as a
function of $\left<\rho \right>/\rho_{\rm crit}$ for haloes averaged
over mass bins in the range $(0.5-2) M_{\rm char}$, where
$M_{\rm char}$, listed in Table~\ref{tab:tab1}, is the typical mass of
equilibrium haloes at each level.  The white curves show median 
values and the surrounding shaded regions the {\em rms} scatter. Even
though we focus on underdense regions, the density around
haloes of mass below $10^{10} {\rm M}_{\odot}$ is centred just below the
mean density and that around more massive haloes is centred just above
the mean density. Furthermore, the value of the environment density
spans at least an order of magnitude, two orders of magnitude in the
case of smaller mass haloes. Concentrations show no monotonic trend
over this range, suggesting that the
concentration-mass relation of Fig~\ref{fig:Fig3} is representative of
the halo population as a whole.

\subsection{Impact of the free-streaming cutoff on halo structure.}
\label{freestream}

We can assess the effect of the free-streaming cutoff on the internal
structure of individual haloes by comparing levels 7 and
7c. We do this in Extended Data Fig.~\ref{fig:FigA5}, which shows density profiles for matched halo pairs in the two simulations. The
haloes are matched by mass (mass difference less than 10 percent) and
separation (offset less than 10 percent of the radius of the high
resolution region). Matched pairs were stacked in four different bins of L7 mass:
$m_{200}= 5\times 10^{-5}$; $10^{-4}$; $5 \times 10^{-4}$ and $10^{-3}\,{\rm M}_\odot$. The numbers of halo pairs in these bins are 152,
132, 40 and 24, respectively. 

The effect of the free-streaming cutoff is to reduce the inner density
(with a corresponding slight increase in the outer density) by an amount that 
grows as the free-streaming mass is approached. The net result is a progressive
reduction in the concentration of haloes with decreasing mass, as may also
be seen by comparing  concentrations for haloes in levels L7 and
L7c in Fig.~\ref{fig:Fig3}. This effect reflects the later
formation of haloes in L7c relative to their
counterparts in L7.

\subsection{Impact of the concentration-mass relation on annihilation luminosities.}
\label{annihilation}

The annihilation luminosity from a dark matter halo scales as the
square of the local dark matter density integrated over its
volume. For smooth profiles of the kind we fit to our simulation,
almost all the luminosity comes from well inside the characteristic
radius ($r_s$ or $r_{-2}$) and scales with the characteristic
parameters of the halo as $V_{\rm max}^4/r_{\rm max}$ where
$V_{\rm max}$ and $r_{\rm max}$ are the maximum circular velocity of
the halo and the radius at which this is attained. The luminosity per
unit mass of a halo thus scales as
$V_{\rm max}^4/(r_{\rm max}M_{200})$ which, at given redshift, depends
only on halo concentration, $c=r_{-2}/r_{200}$, for Einasto halos with
constant $\alpha$ ($=0.16$ in our case).

For the well resolved haloes at each level of our simulation, we can measure $V_{\rm max}$ and $r_{\rm max}$ directly and so estimate their light-to mass ratios. Averaging over all haloes of given mass and multiplying by their contribution to the mass density of the universe according to some halo mass function (for illustration we here use the analytic Sheth-Tormen function\cite{Sheth_02}) we can construct Extended Data Fig.~\ref{fig:FigA6} which shows how the total annihilation luminosity of the present universe is distributed over halo mass. Remarkably, we find the contribution per unit logarithmic halo mass interval to be almost constant over the range $-3\leq \ln M_{200}/{\rm M}_\odot\leq 11$. This is considerably less weighted towards low-mass haloes than estimates based on previously published extrapolations of the concentration-mass relation, and the total luminosity density is lower by factors ranging up to $10^3$. Thus, the significance of very small structures for annihilation luminosities has been overestimated in the past, often by substantial factors. Note that this observation is likely to apply to substructure within haloes as well as to emission from the main (smooth) halo, even though we have not considered 
    such subhalo contributions here. 
    
\clearpage

\section*{Additional references}
\vspace{0.5cm}

\bibliographystyle{naturemag}

\begin{thebibliography}{10}
        \expandafter\ifx\csname url\endcsname\relax
          \def\url#1{\texttt{#1}}\fi
        \expandafter\ifx\csname urlprefix\endcsname\relax\def\urlprefix{URL }\fi
        \providecommand{\bibinfo}[2]{#2}
        \providecommand{\eprint}[2][]{\url{#2}}

\bibitem{Bluementhal_84}
\bibinfo{author}{{Blumenthal}, G.~R.}, \bibinfo{author}{{Faber}, S.~M.}, \bibinfo{author}{{Primack}, J.~R.} \& \bibinfo{author}{{Rees}, M.~J.}
\newblock \bibinfo{title}{{Formation of galaxies and large-scale structure with cold dark
  matter}}.
\newblock \emph{\bibinfo{journal}{\nat}} \textbf{\bibinfo{volume}{311}},
  \bibinfo{pages}{517--525} (\bibinfo{year}{1984}).

\bibitem{White_91}
\bibinfo{author}{{White}, S. ~D.~M.} \& \bibinfo{author}{{Frenk}, C. ~S.}
\newblock \newblock \bibinfo{title}{{Galaxy Formation through Hierarchical Clustering}}.
\newblock \emph{\bibinfo{journal}{\apj}} \textbf{\bibinfo{volume}{379}},
  \bibinfo{pages}{52} (\bibinfo{year}{1991}).

\bibitem{Bertone2005}
\bibinfo{author}{{Bertone}, G.}, \bibinfo{author}{{Hooper}, D.} \&
  \bibinfo{author}{{Silk}, J.}
\newblock \bibinfo{title}{{Particle dark matter: evidence, candidates and
  constraints}}.
\newblock \emph{\bibinfo{journal}{\physrep}} \textbf{\bibinfo{volume}{405}},
  \bibinfo{pages}{279--390} (\bibinfo{year}{2005}).

\bibitem{NFW_97}
\bibinfo{author}{{Navarro}, J.~F.}, \bibinfo{author}{{Frenk}, C.~S.} \&
  \bibinfo{author}{{White}, S.~D.~M.}
\newblock \bibinfo{title}{{A Universal Density Profile from Hierarchical Clustering}}.
\newblock \emph{\bibinfo{journal}{\apj}} \textbf{\bibinfo{volume}{490}},
  \bibinfo{pages}{493} (\bibinfo{year}{1997}).

\bibitem{Navarro_04}
\bibinfo{author}{{Navarro}, J.~F.} \emph{et~al.}
\newblock \bibinfo{title}{{The inner structure of {\ensuremath{\Lambda}}CDM
  haloes - III. Universality and asymptotic slopes}}.
\newblock \emph{\bibinfo{journal}{\mnras}} \textbf{\bibinfo{volume}{349}},
  \bibinfo{pages}{1039--1051} (\bibinfo{year}{2004}).
 


\bibitem{Springel_05}
\bibinfo{author}{{Springel}, V.} \emph{et~al.}
\newblock \bibinfo{title}{{Simulations of the formation, evolution and
  clustering of galaxies and quasars}}.
\newblock \emph{\bibinfo{journal}{\nat}} \textbf{\bibinfo{volume}{435}},
  \bibinfo{pages}{629--636} (\bibinfo{year}{2005}).
 

\bibitem{NFW_96}
\bibinfo{author}{{Navarro}, J.~F.}, \bibinfo{author}{{Frenk}, C.~S.} \&
  \bibinfo{author}{{White}, S.~D.~M.}
\newblock \bibinfo{title}{{The Structure of Cold Dark Matter Halos}}.
\newblock \emph{\bibinfo{journal}{\apj}} \textbf{\bibinfo{volume}{462}},
  \bibinfo{pages}{563} (\bibinfo{year}{1996}).
 

\bibitem{Einasto_65}
\bibinfo{author}{{Einasto}, J.}
\newblock \bibinfo{title}{{On the Construction of a Composite Model for the
  Galaxy and on the Determination of the System of Galactic Parameters}}.
\newblock \emph{\bibinfo{journal}{Trudy Astrofizicheskogo Instituta Alma-Ata}}
  \textbf{\bibinfo{volume}{5}}, \bibinfo{pages}{87--100}
  (\bibinfo{year}{1965}).

\bibitem{Ludlow_14}
\bibinfo{author}{{Ludlow}, A.~D.} \emph{et~al.}
\newblock \bibinfo{title}{{The mass-concentration-redshift relation of cold
  dark matter haloes}}.
\newblock \emph{\bibinfo{journal}{\mnras}} \textbf{\bibinfo{volume}{441}},
  \bibinfo{pages}{378--388} (\bibinfo{year}{2014}).
 

\bibitem{Ludlow_16}
\bibinfo{author}{{Ludlow}, A.~D.} \emph{et~al.}
\newblock \bibinfo{title}{{The mass-concentration-redshift relation of cold and
  warm dark matter haloes}}.
\newblock \emph{\bibinfo{journal}{\mnras}} \textbf{\bibinfo{volume}{460}},
  \bibinfo{pages}{1214--1232} (\bibinfo{year}{2016}).
 

\bibitem{Springel_08a}
\bibinfo{author}{Springel, V.} \emph{et~al.}
\newblock \bibinfo{title}{The aquarius project: the subhaloes of galactic
  haloes}.
\newblock \emph{\bibinfo{journal}{MNRAS}} \textbf{\bibinfo{volume}{391}},
  \bibinfo{pages}{1685--1711} (\bibinfo{year}{2008}).

\bibitem{Stuecker2018}
\bibinfo{author}{{St{\"u}cker}, J.}, \bibinfo{author}{{Busch}, P.} \&
  \bibinfo{author}{{White}, S. D.~M.}
\newblock \bibinfo{title}{{The median density of the Universe}}.
\newblock \emph{\bibinfo{journal}{\mnras}} \textbf{\bibinfo{volume}{477}},
  \bibinfo{pages}{3230--3246} (\bibinfo{year}{2018}).
 

\bibitem{Diemand2005}
\bibinfo{author}{{Diemand}, J.}, \bibinfo{author}{{Moore}, B.} \&
  \bibinfo{author}{{Stadel}, J.}
\newblock \bibinfo{title}{{Earth-mass dark-matter haloes as the first
  structures in the early Universe}}.
\newblock \emph{\bibinfo{journal}{\nat}} \textbf{\bibinfo{volume}{433}},
  \bibinfo{pages}{389--391} (\bibinfo{year}{2005}).
 

\bibitem{Ishiyama2010}
\bibinfo{author}{{Ishiyama}, T.}, \bibinfo{author}{{Makino}, J.} \&
  \bibinfo{author}{{Ebisuzaki}, T.}
\newblock \bibinfo{title}{{Gamma-ray Signal from Earth-mass Dark Matter
  Microhalos}}.
\newblock \emph{\bibinfo{journal}{\apjl}} \textbf{\bibinfo{volume}{723}},
  \bibinfo{pages}{L195--L200} (\bibinfo{year}{2010}).
 

\bibitem{Anderhalden2013}
\bibinfo{author}{{Anderhalden}, D.} \& \bibinfo{author}{{Diemand}, J.}
\newblock \bibinfo{title}{{Density profiles of CDM microhalos and their
  implications for annihilation boost factors}}.
\newblock \emph{\bibinfo{journal}{JCAP}} \textbf{\bibinfo{volume}{2013}},
  \bibinfo{pages}{009} (\bibinfo{year}{2013}).
 

\bibitem{Angulo2017}
\bibinfo{author}{{Angulo}, R.~E.}, \bibinfo{author}{{Hahn}, O.},
  \bibinfo{author}{{Ludlow}, A.~D.} \& \bibinfo{author}{{Bonoli}, S.}
\newblock \bibinfo{title}{{Earth-mass haloes and the emergence of NFW density
  profiles}}.
\newblock \emph{\bibinfo{journal}{\mnras}} \textbf{\bibinfo{volume}{471}},
  \bibinfo{pages}{4687--4701} (\bibinfo{year}{2017}).
 

\bibitem{Tseliakhovich2010}
\bibinfo{author}{{Tseliakhovich}, D.} \& \bibinfo{author}{{Hirata}, C.}
\newblock \bibinfo{title}{{Relative velocity of dark matter and baryonic fluids
  and the formation of the first structures}}.
\newblock \emph{\bibinfo{journal}{PRD}} \textbf{\bibinfo{volume}{82}},
  \bibinfo{pages}{083520} (\bibinfo{year}{2010}).
 

\bibitem{Neto_07}
{Neto}, A.~F. \emph{et~al.}
\newblock {The statistics of {\ensuremath{\Lambda}} CDM halo concentrations}.
\newblock \emph{\bibinfo{journal}{\mnras}} \textbf{\bibinfo{volume}{381}},
  \bibinfo{pages}{1450--1462} (\bibinfo{year}{2007}).

\bibitem{Sanchez-Conde_14}
{S{\'a}nchez-Conde}, M.~A. \&  {Prada}, F.
\newblock {The flattening of the concentration-mass relation towards low halo
  masses and its implications for the annihilation signal boost}.
\newblock \emph{\bibinfo{journal}{\mnras}} \textbf{\bibinfo{volume}{442}},
  \bibinfo{pages}{2271--2277} (\bibinfo{year}{2014}).


\bibitem{Dutton_14}
{Dutton}, A.~A. \& {Macci{\`o}}, A.~V. 
\newblock {Cold dark matter haloes in the Planck era: evolution of structural
  parameters for Einasto and NFW profiles}.
\newblock \emph{\bibinfo{journal}{\mnras}} \textbf{\bibinfo{volume}{441}},
  \bibinfo{pages}{3359--3374} (\bibinfo{year}{2014}).

\bibitem{Diemer_19}
{Diemer}, B. \& {Joyce}, M.
\newblock {An Accurate Physical Model for Halo Concentrations}.
\newblock \emph{\bibinfo{journal}{\apj}} \textbf{\bibinfo{volume}{871}},
  \bibinfo{pages}{168} (\bibinfo{year}{2019}).

\end{thebibliography}

\begin{thebibliography}{10}
        \expandafter\ifx\csname url\endcsname\relax
          \def\url#1{\texttt{#1}}\fi
        \expandafter\ifx\csname urlprefix\endcsname\relax\def\urlprefix{URL }\fi
        \providecommand{\bibinfo}[2]{#2}
        \providecommand{\eprint}[2][]{\url{#2}}
 
 \setcounter{enumiv}{21}


\bibitem{Planck_14}
\bibinfo{author}{{Planck Collaboration}} \emph{et~al.}
\newblock \bibinfo{title}{{Planck 2013 results. I. Overview of products and
  scientific results}}.
\newblock \emph{\bibinfo{journal}{\aap}} \textbf{\bibinfo{volume}{571}},
  \bibinfo{pages}{A1} (\bibinfo{year}{2014}).
 

\bibitem{Schaye_15}
\bibinfo{author}{{Schaye}, J.} \emph{et~al.}
\newblock \bibinfo{title}{{The EAGLE project: simulating the evolution and
  assembly of galaxies and their environments}}.
\newblock \emph{\bibinfo{journal}{\mnras}} \textbf{\bibinfo{volume}{446}},
  \bibinfo{pages}{521--554} (\bibinfo{year}{2015}).
 

\bibitem{Lewis_00}
\bibinfo{author}{{Lewis}, A.}, \bibinfo{author}{{Challinor}, A.} \&
  \bibinfo{author}{{Lasenby}, A.}
\newblock \bibinfo{title}{{Efficient Computation of Cosmic Microwave Background
  Anisotropies in Closed Friedmann-Robertson-Walker Models}}.
\newblock \emph{\bibinfo{journal}{\apj}} \textbf{\bibinfo{volume}{538}},
  \bibinfo{pages}{473--476} (\bibinfo{year}{2000}).
 

\bibitem{BBKS}
\bibinfo{author}{{Bardeen}, J.~M.}, \bibinfo{author}{{Bond}, J.~R.},
  \bibinfo{author}{{Kaiser}, N.} \& \bibinfo{author}{{Szalay}, A.~S.}
\newblock \bibinfo{title}{{The Statistics of Peaks of Gaussian Random Fields}}.
\newblock \emph{\bibinfo{journal}{\apj}} \textbf{\bibinfo{volume}{304}},
  \bibinfo{pages}{15} (\bibinfo{year}{1986}).

\bibitem{Springel_08b}
\bibinfo{author}{{Springel}, V.} \emph{et~al.}
\newblock \bibinfo{title}{{Prospects for detecting supersymmetric dark matter
  in the Galactic halo}}.
\newblock \emph{\bibinfo{journal}{\nat}} \textbf{\bibinfo{volume}{456}},
  \bibinfo{pages}{73--76} (\bibinfo{year}{2008}).
 

\bibitem{Green_04}
\bibinfo{author}{{Green}, A.~M.}, \bibinfo{author}{{Hofmann}, S.} \&
  \bibinfo{author}{{Schwarz}, D.~J.}
\newblock \bibinfo{title}{{The power spectrum of SUSY-CDM on subgalactic
  scales}}.
\newblock \emph{\bibinfo{journal}{\mnras}} \textbf{\bibinfo{volume}{353}},
  \bibinfo{pages}{L23--L27} (\bibinfo{year}{2004}).
 
 \bibitem{Martinez_09}
{Martinez}, G.~D.  \emph{et~al.}
\newblock {Indirect Dark Matter detection from Dwarf satellites: joint
  expectations from astrophysics and supersymmetry}.
\newblock {\em \jcap} {\bf 06}, 014 (2009).
 
\bibitem{Bertone_18}
{Bertone}, G. \& {Tait}, T.~M.~P. 
\newblock {A new era in the search for dark matter}.
 \newblock \emph{\bibinfo{journal}{\nat}} \textbf{\bibinfo{volume}{562}},
  \bibinfo{pages}{51--56} (\bibinfo{year}{2018}).

 
 
\bibitem{Jenkins2010}
\bibinfo{author}{{Jenkins}, A.}
\newblock \bibinfo{title}{{Second-order Lagrangian perturbation theory initial
  conditions for resimulations}}.
\newblock \emph{\bibinfo{journal}{\mnras}} \textbf{\bibinfo{volume}{403}},
  \bibinfo{pages}{1859--1872} (\bibinfo{year}{2010}).
 

\bibitem{Jenkins2013}
\bibinfo{author}{{Jenkins}, A.}
\newblock \bibinfo{title}{{A new way of setting the phases for cosmological
  multiscale Gaussian initial conditions}}.
\newblock \emph{\bibinfo{journal}{\mnras}} \textbf{\bibinfo{volume}{434}},
  \bibinfo{pages}{2094--2120} (\bibinfo{year}{2013}).
 

\bibitem{JenkinsBooth2013}
\bibinfo{author}{{Jenkins}, A.} \& \bibinfo{author}{{Booth}, S.}
\newblock \bibinfo{title}{{Panphasia: a user guide}}.
\newblock \emph{\bibinfo{journal}{arXiv e-prints}}
  \bibinfo{pages}{arXiv:1306.5771} (\bibinfo{year}{2013}).
 

\bibitem{Gadget2}
\bibinfo{author}{{Springel}, V.}
\newblock \bibinfo{title}{{The cosmological simulation code GADGET-2}}.
\newblock \emph{\bibinfo{journal}{\mnras}} \textbf{\bibinfo{volume}{364}},
  \bibinfo{pages}{1105--1134} (\bibinfo{year}{2005}).
 

\bibitem{Barnes1986}
\bibinfo{author}{{Barnes}, J.} \& \bibinfo{author}{{Hut}, P.}
\newblock \bibinfo{title}{{A hierarchical O(N log N) force-calculation
  algorithm}}.
\newblock \emph{\bibinfo{journal}{\nat}} \textbf{\bibinfo{volume}{324}},
  \bibinfo{pages}{446--449} (\bibinfo{year}{1986}).

\bibitem{Dehnen2000}
\bibinfo{author}{{Dehnen}, W.}
\newblock \bibinfo{title}{{A Very Fast and Momentum-conserving Tree Code}}.
\newblock \emph{\bibinfo{journal}{\apjl}} \textbf{\bibinfo{volume}{536}},
  \bibinfo{pages}{L39--L42} (\bibinfo{year}{2000}).
 

\bibitem{Pelupessy2012}
\bibinfo{author}{{Pelupessy}, F.~I.}, \bibinfo{author}{{J{\"a}nes}, J.} \&
  \bibinfo{author}{{Portegies Zwart}, S.}
\newblock \bibinfo{title}{{N-body integrators with individual time steps from
  Hierarchical splitting}}.
\newblock \emph{\bibinfo{journal}{\na}} \textbf{\bibinfo{volume}{17}},
  \bibinfo{pages}{711--719} (\bibinfo{year}{2012}).
 

\bibitem{Wang_07}
\bibinfo{author}{{Wang}, J.} \& \bibinfo{author}{{White}, S. D.~M.}
\newblock \bibinfo{title}{{Discreteness effects in simulations of hot/warm dark
  matter}}.
\newblock \emph{\bibinfo{journal}{\mnras}} \textbf{\bibinfo{volume}{380}},
  \bibinfo{pages}{93--103} (\bibinfo{year}{2007}).
 

\bibitem{Power_03}
\bibinfo{author}{Power, C.} \emph{et~al.}
\newblock \bibinfo{title}{The inner structure of {$\Lambda$}cdm haloes - i. a
  numerical convergence study}.
\newblock \emph{\bibinfo{journal}{MNRAS}} \textbf{\bibinfo{volume}{338}},
  \bibinfo{pages}{14--34} (\bibinfo{year}{2003}).

\bibitem{Gao_08}
\bibinfo{author}{{Gao}, L.} \emph{et~al.}
\newblock \bibinfo{title}{{The redshift dependence of the structure of massive
  {\ensuremath{\Lambda}} cold dark matter haloes}}.
\newblock \emph{\bibinfo{journal}{\mnras}} \textbf{\bibinfo{volume}{387}},
  \bibinfo{pages}{536--544} (\bibinfo{year}{2008}).
 

\bibitem{Lavalle_08}
\bibinfo{author}{{Lavalle}, J.}, \bibinfo{author}{{Yuan}, Q.},
  \bibinfo{author}{{Maurin}, D.} \& \bibinfo{author}{{Bi}, X.~J.}
\newblock \bibinfo{title}{{Full calculation of clumpiness boost factors for
  antimatter cosmic rays in the light of {\ensuremath{\Lambda}}CDM N-body
  simulation results. Abandoning hope in clumpiness enhancement?}}
\newblock \emph{\bibinfo{journal}{\aap}} \textbf{\bibinfo{volume}{479}},
  \bibinfo{pages}{427--452} (\bibinfo{year}{2008}).
 
           
\bibitem{Sheth_02}
{{Sheth}, R.K} \& {{Tormen}, G.}
\newblock {An excursion set model of hierarchical clustering: ellipsoidal
  collapse and the moving barrier}.
\newblock \emph{\bibinfo{journal}{MNRAS}} \textbf{\bibinfo{volume}{329}},
  \bibinfo{pages}{61--75} (\bibinfo{year}{2002}).
  
             
\end{thebibliography}

\expandafter\ifx\csname url\endcsname\relax
  \def\url#1{\texttt{#1}}\fi
\expandafter\ifx\csname urlprefix\endcsname\relax\def\urlprefix{URL }\fi
\providecommand{\bibinfo}[2]{#2}
\providecommand{\eprint}[2][]{\url{#2}}
\setlength{\itemsep}{0ex}

\begin{addendum}
 \item[Competing Interests] The authors declare that they have no
competing financial interests.

\item[Data Availability] The data generated, analysed and presented in this study are available from the corresponding authors on request, though the requester will be responsible for providing the very considerable resources needed for transferring and storing these data.
  
\item[Code Availability] The parent code GADGET-2 has been publicly available for some time. It is expected that most of the extensions and modifications made in order to meet the extreme requirements of this project will be made available in the future release of GADGET-4;  those interested can contact VS for further information. The very large white noise field, Panphasia, used to code our initial conditions, is also publicly available; those interested can contact ARJ.

\item[Additional information] None
\item[Correspondence and requests for materials] should be addressed
  to JW and SB.
\end{addendum}
\clearpage
\newpage

\renewcommand*{\figurename}{Extended Data Figure}
\setcounter{figure}{0}

\renewenvironment{figure}{\let\caption\edfigcaption}{}

\begin{figure}
\begin{center}
\vspace{-3cm}
\includegraphics[width=0.85\textwidth,angle=0,trim=0cm 3cm 1cm 4cm,clip=true]{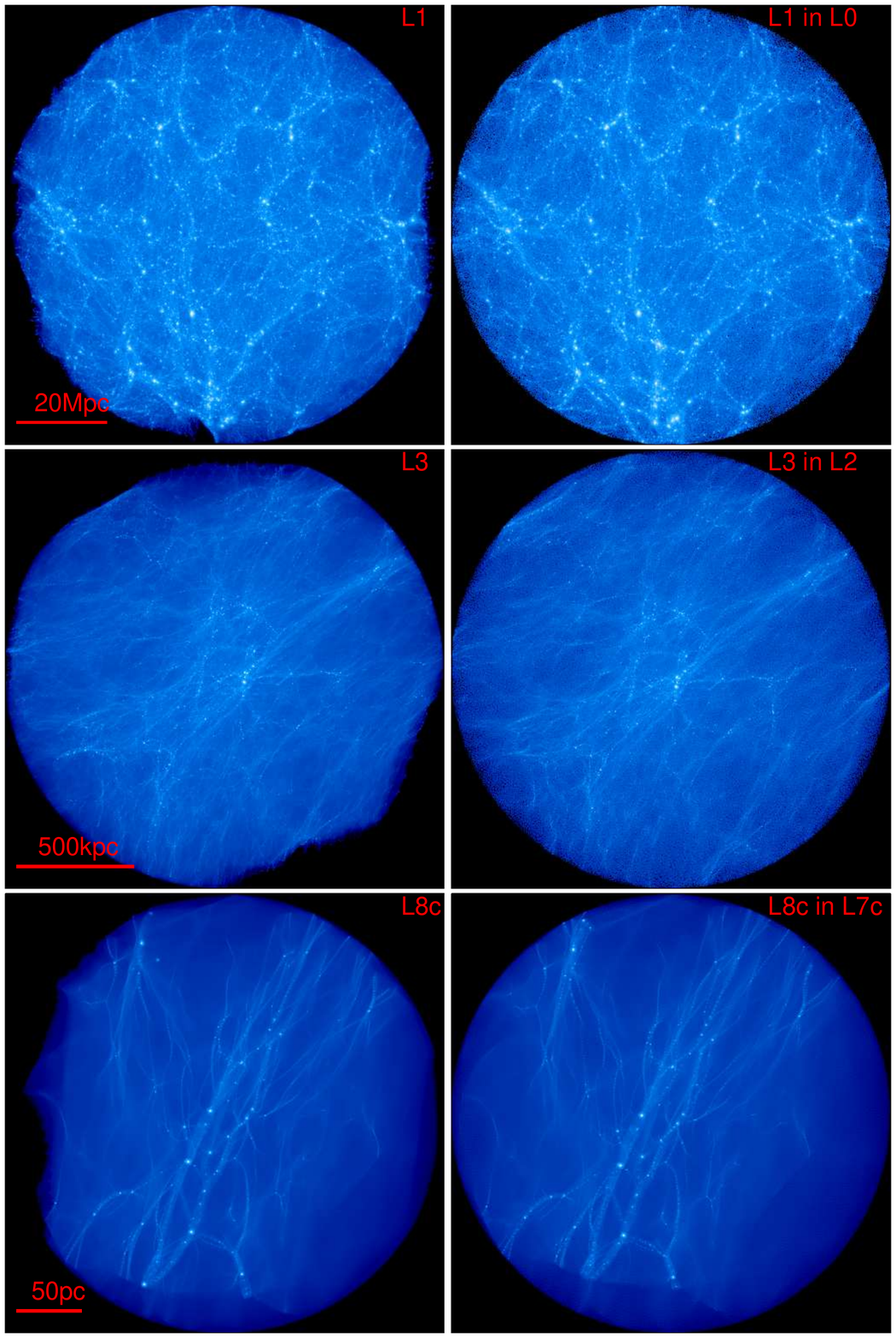}
\end{center}
\caption{\linespread{1.3}\selectfont{}\rm { \bf Projected density maps
    for different zoom levels.} L1, L3 and L8c (left) are
    compared with maps of the same regions in their parent levels
    L0, L2, and L7c, respectively (right). The regions shown are
  the largest spheres that fit almost entirely within the high
  resolution region of the higher level. Only high resolution
  particles are used to make the images.}
\label{fig:FigA1}
\end{figure}

\begin{figure}
\begin{center}
\includegraphics[width=0.7\textwidth,trim=2.5cm 2cm 0cm 6cm,clip=true] {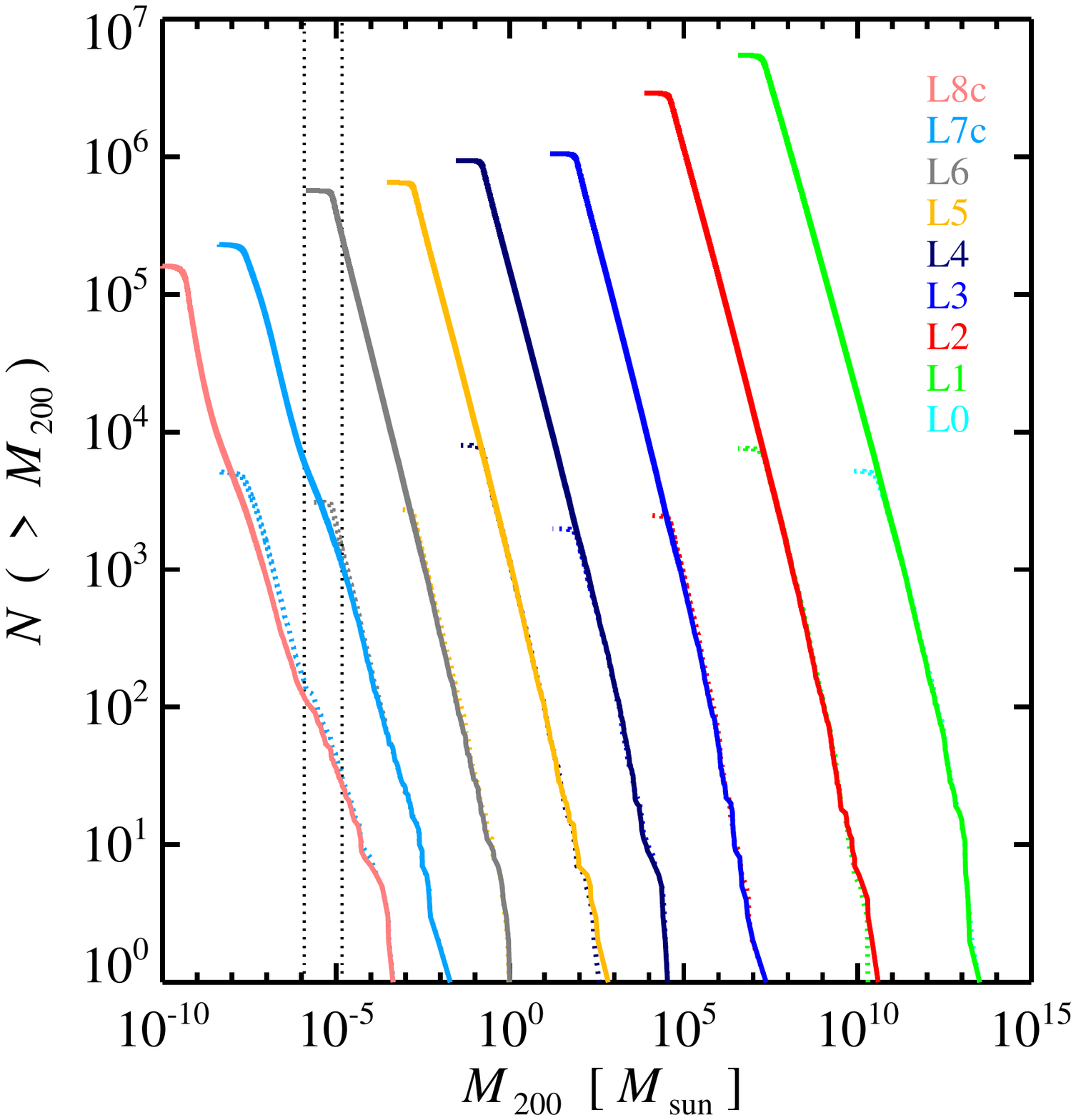}
\end{center}
\caption{\linespread{1.3}\rm { \bf The cumulative halo
    number as a function of mass, 
    $M_{200}$, in the high-resolution region of each zoom level 
    compared to that in the same region of the parent level.}
  Different colours denote different levels as indicated in the
  legend. Results from the parent levels are shown as dotted
  curves. The two vertical black dotted lines indicate the upper mass
  limit for spurious haloes in L7c and L8c, calculated as described Wang et al.\cite{Wang_07}. Note the excellent agreement between the solid and dotted
  curves above the resolution limit of the latter (and above the L7c
  mass limit for spurious haloes in the case of L7c versus L8c).  }
\label{fig:FigA2}
\end{figure}

\begin{figure}
\begin{center}
\includegraphics[width=0.6\textwidth,angle=0,trim=0cm 4cm 1cm 6cm,clip=true]{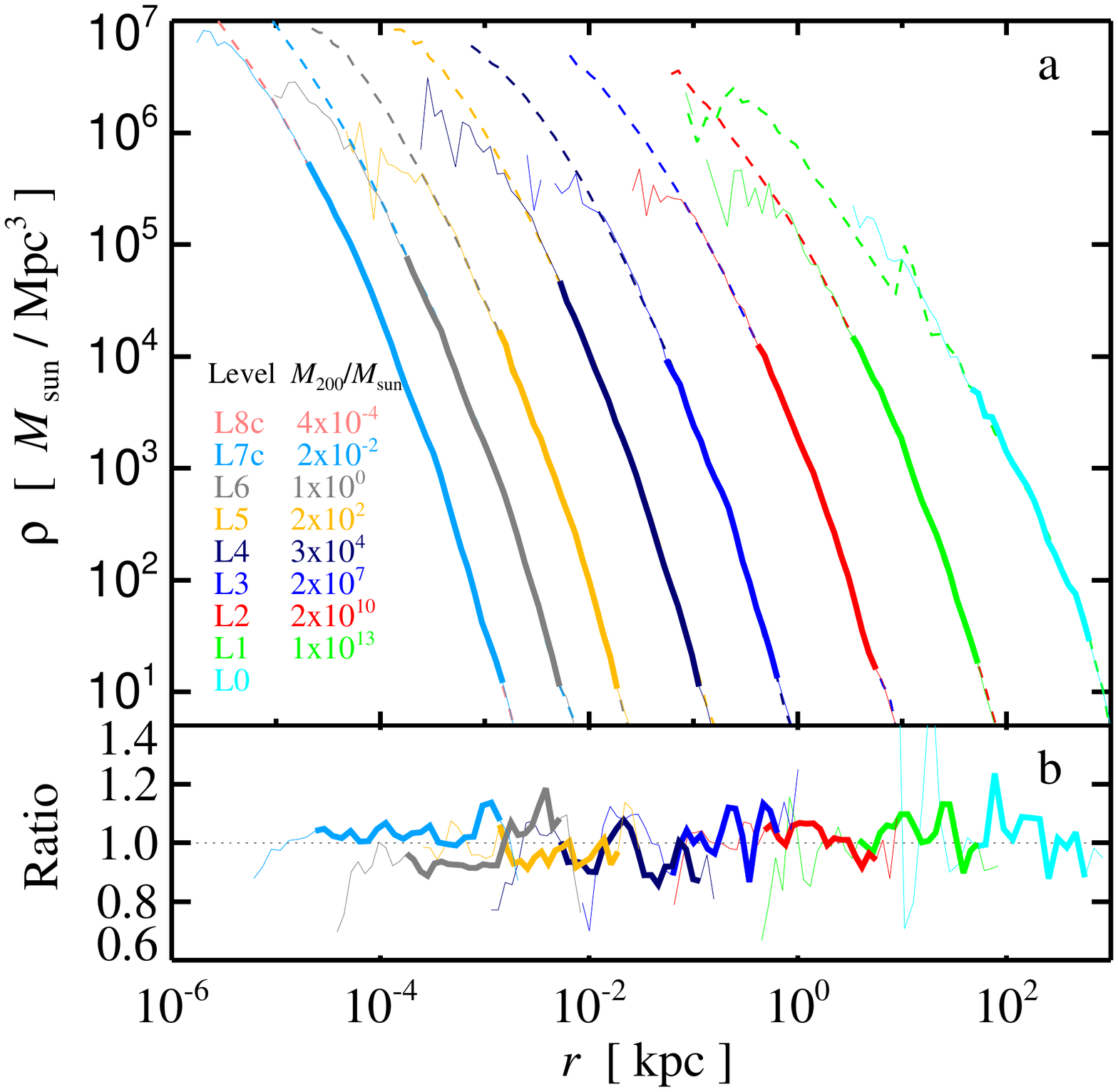}
\end{center}
\caption{\linespread{1.3}\rm {\bf Comparison of the density
    profile of a massive halo at each level with that of its
    counterpart in the parent level.} {\it Panel a:} the density profile
  of one of the most massive haloes in the high-resolution region of
  each zoom level is compared to that of the same halo at the parent
  level.  Results from different levels are shown with different
  colours, as indicated by the legend, which also gives the masses of
  the haloes concerned. Higher resolution profiles are shown as dashed
  curves, while those from the parent levels are shown as solid
  curves. The range between the convergence radius and $r_{200}$ is
  plotted as a thick line in the lower resolution case. {\it Panel b:}
  the ratio of the density profiles of each pair in the upper
  panel. Again, results in the range between the convergence radius in
  the lower resolution case and $r_{200}$ are shown as thick
  lines. Note the excellent convergence between simulation pairs over
  this radial range, which typically differ in mass resolution by a
  factor of several hundred.}
\label{fig:FigA3}
\end{figure}

\begin{figure}
\begin{center}
\includegraphics[width=0.75\textwidth,angle=0,trim=1.8cm 2.5cm 0cm 9cm,clip=true]{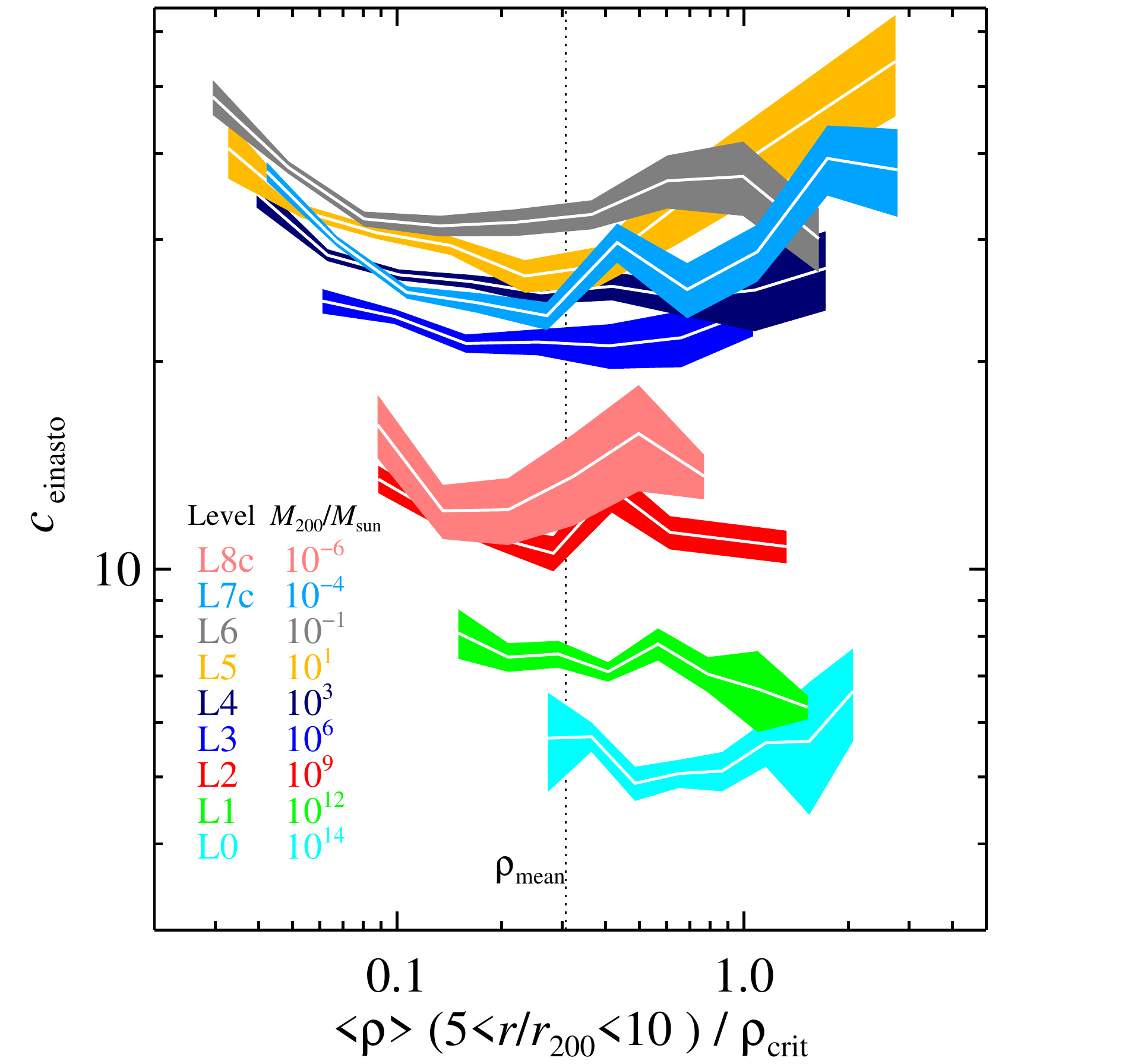}
\end{center}
\caption{\linespread{1.3}\rm {\bf Dependence of halo
    concentration on local environment in the high-resolution region
    at each zoom level.} Results are shown for haloes in the mass
  range $[0.5, 2]M_{\rm char}$; the legend gives the
  characteristic mass, $M_{\rm char}$, for each level and also defines
  the colour key. Each white curve gives the median concentration for
  the best-fit Einasto profile, while the surrounding coloured region
  gives the {\it rms} scatter.  Local environment density is defined
  here as the mean in a thick spherical shell, $5 <r/r_{200}<10$,
  surrounding each halo, and is given in units of the critical
  density. All haloes are used for this plot. A vertical line shows
  the cosmic mean density. Note that although concentration depends
  significantly on mass, any dependence on local environment density
  is weak.}
\label{fig:FigA4}
\end{figure}

\begin{figure}
\begin{center}
\vspace{-1cm}
\includegraphics[width=0.8\textwidth,height=13cm, angle=0,trim=2cm 2cm 0cm 8cm,clip=true]{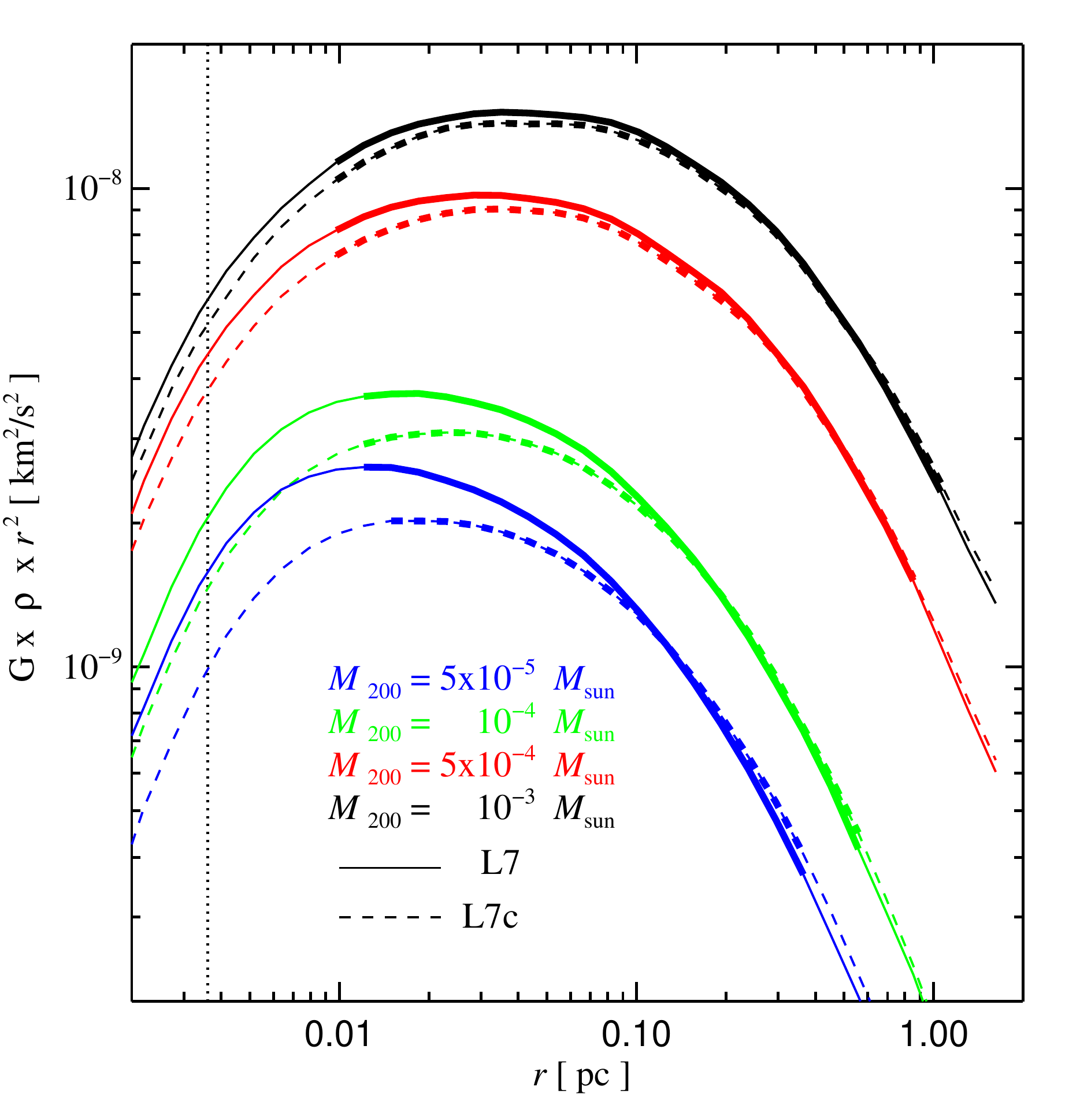}
\end{center}
\caption{\linespread{1.3}\rm {\bf Stacked density profiles of matched haloes in the L7 (solid) and L7c (dashed) simulations.} The densities are multipled
  by $r^2$ to increase the dynamical range of the figure. Different
  colours correspond to different mass bins with central values quoted
  in the legend. The profiles are shown as thick lines over the range
  where they are most reliable, between the convergence radius,
  $r_{\rm conv}$, and $r_{200}$.  The vertical dotted line indicates
  the softening length in the high-resolution region at this
  level. The effect of the free-streaming cutoff is to reduce the
  density in the inner parts, and therefore the concentration, by an
  increasing amount as the halo mass approaches the free-streaming mass.}
\label{fig:FigA5}
\end{figure}

\begin{figure}
\begin{center}
\vspace{-1cm}
\includegraphics[width=0.8\textwidth,height=11cm, angle=0,trim=0cm 5cm 0cm 5cm,clip=true]{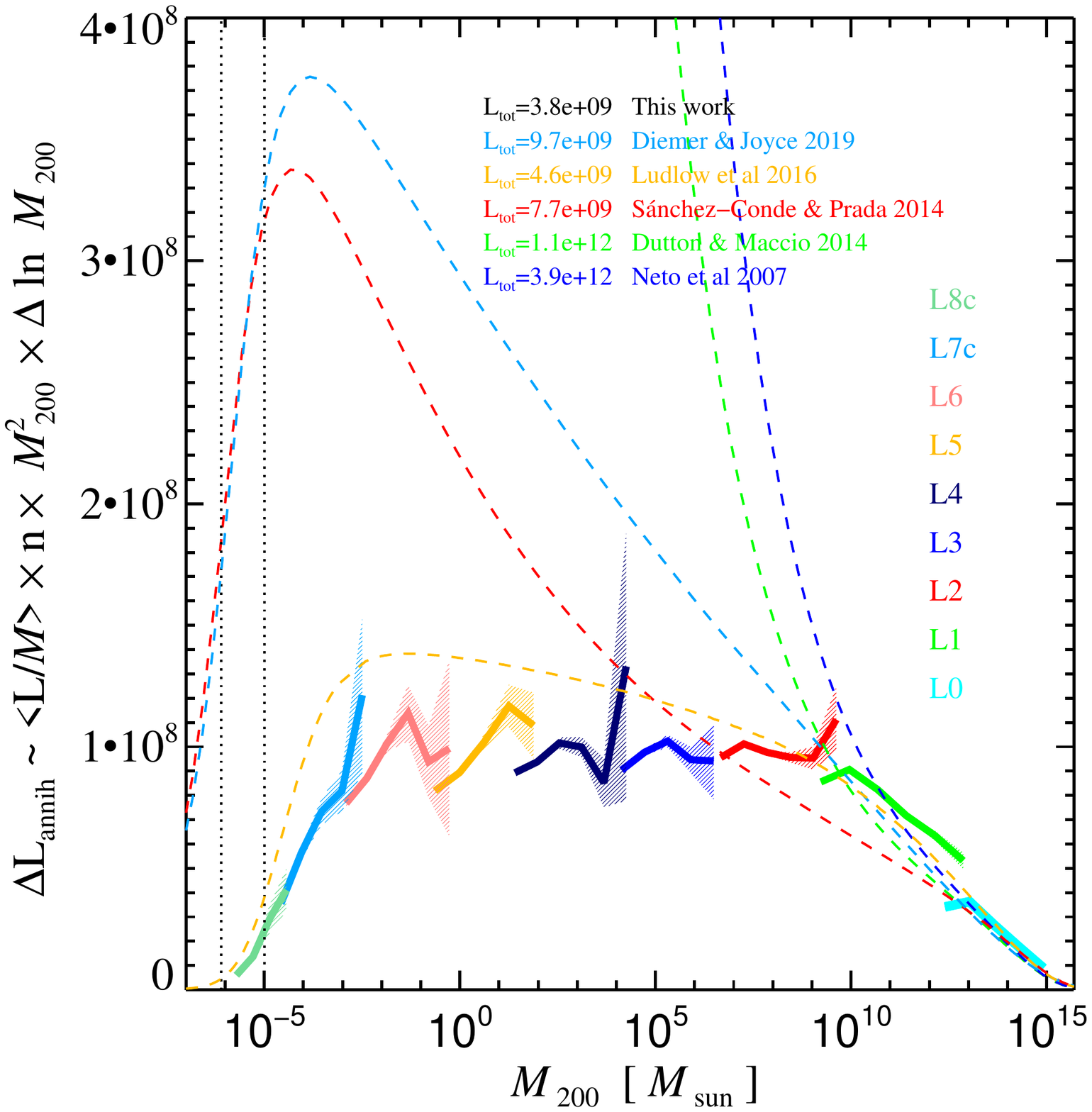}
\end{center}
\caption{\linespread{1.3}\rm \bf The mean luminosity density produced by dark matter annihilation in today's universe as a function of halo mass.
Solid coloured lines indicate estimates obtained by multiplying the mean $\langle L/M\rangle$ obtained from well resolved, simulated haloes by the halo number density predicted for each bin by the Sheth-Tormen halo mass function.\cite{Sheth_02} Shaded regions indicate the estimated 1$\sigma$ uncertainty. Dashed curves indicate results found if we instead use $\langle L/M\rangle$ values predicted by previously published mass-concentration relations. Corresponding predictions for the total integrated luminosity density are given in the legend by values preceding each model's name. Note that the units of annihilation luminosity are arbitrary, so only ratios of values are significant.}
\label{fig:FigA6}
\end{figure}

\renewcommand{\tablename}{Extended Data Table}
\setcounter{table}{0}

\begin{table}
\centering
\caption{\linespread{1.3}\selectfont{} \textbf{Parameters of the simulation levels.} {\em Column 1:}
  name of the level; {\em Column 2:} 
  $R_{\rm high}$, the radius of the high-resolution region; {\em Column 3:}
  $n_{\rm p}$, the total
  number of high-resolution particles; {\em Column 4:}
  $\epsilon$, the softening
  length of the high-resolution particles; {\em Column 5:}
  $m_{\rm p}$, the mass of the high-resolution particles; 
  {\em Column 6:} $\sigma (M_{\rm tot},z=0)$, the {\it rms} linear overdensity, extrapolated to $z=0$, within spheres that on average contain mass $M_{\rm tot}=n_p m_p$. {\em Column 7:} ${\left<\rho\right>/\rho_{\rm mean}}$, the mean mass density at $z=0$ in the high
  resolution region in units of the cosmic mean; {\em Column 8:}
  $M_{\rm char}$, the typical mass of the equilibrium haloes for which profiles
  were stacked in Fig.~2. {\em Column 9:} $N_{\rm char}$, the number of
  haloes in the  mass bin $[0.8\,M_{\rm char},\, 1.2\,M_{\rm char}$] used in the stacks.  {\em Column 10:}
  $z_{\rm  form}$, the median formation redshift of equilibrium haloes of the characteristic mass. {\em Column 11:}
  $f_{\rm   vir}$, the fraction of haloes with more than 3000 particles that are in equilibrium
  according to the criteria given in the text.  }
  
\label{tab:tab1}
\begin{center}

\scalebox{0.8}{
\begin{tabular}{ccccccccccc} \hline
{\rm level} & $R_{\rm high}\;[ {\rm Mpc}]$  &  $n_{\rm p}$ & $\epsilon\;[ {\rm  kpc}]$ & $m_{\rm  p}\;[M_{\odot}]$ & $\sigma (M_{\rm tot},z=0)$  & ${\left<\rho\right>/\rho_{\rm mean}}$ & $M_{\rm char}\; [M_{\odot}]$ &  $N_{\rm char}$  & $z_{\rm form}$ & $f_{\rm vir}$ \\ \hline 
{\rm L0}  & $738$ & $1.0 \times 10^{10}$ & $7.4$ &  $1.5\times 10^9$ & & $1.0$ & $10^{14}$ & $127$ & $0.94$ & $0.92$  \\
{\rm L1}  & $52$ & $1.0\times10^{10}$ & $4.4\times 10^{-1}$ &  $7.4 \times  10^5$ & $0.34$  & $0.39$ & $10^{12}$ & $59$ & $1.66$ &$0.91$ \\
{\rm L2}  &  $8.8$ & $5.4\times10^9$ & $5.6\times 10^{-2}$ &  $1.5\times 10^3$& $1.66$ &  $0.082$ & $10^9$ & $29$ &$1.91$ & $0.93$  \\ 
{\rm L3}  & $1.0$ & $1.8\times 10^9$ & $8.3\times 10^{-3}$ &  $2.8$ & $4.22$ &  $0.036$ & $10^6$ & $27$ & $2.61$ & $0.94$  \\
{\rm L4}  & $0.27$& $2.0\times 10^9$ & $1.0\times 10^{-3}$ & $5.5 \times 10^{-3}$ & $6.96$ &  $0.026$ & $10^3$ & $59$ & $4.44$ &$0.94$  \\
{\rm L5}  & $0.035$& $1.5\times 10^9$ & $2.2\times 10^{-4}$ & $5.8\times 10^{-5}$ & $9.36$ &  $0.024$ & $10$ & $30$ & $4.68$  &$0.94$ \\
{\rm L6}  & $0.0066$ &$1.7\times 10^9$ & $3.8\times 10^{-5}$ & $2.6\times10^{-7}$ & $12.12$ &   $0.014$    &
$10^{-1}$ & $35$  & $4.84$ & $0.94$   \\
{\rm L7}  & $0.0011$ & $2.5 \times 10^9$ & $5.3\times 10^{-6}$ & $8.6\times 10^{-10}$  & $15.06$ &  $0.016$ & $10^{-4}$ & $201$ & $5.21$ & $0.96$  \\
{\rm L7c}  &$0.0011$  & $2.5\times 10^9$ & $5.3\times 10^{-6}$ & $8.6\times 10^{-10}$ & $15.06$ & $0.016$ & $10^{-4}$ & $202$ & $ 4.83$ & $0.97$  \\
{\rm L8c}  & $0.00024$ &$1.5 \times 10^9$ & $1.4\times 10^{-6}$ & $1.6\times 10^{-11}$ & $17.60$ &  $0.028$ & $10^{-6}$&  $24$ & $1.96$ & $0.94$  \\
\hline
\end{tabular}}
\end{center}
\end{table}

\end{methods}

\end{document}